%% file: main.tex
\documentclass[10pt,conference]{IEEEtran}
\usepackage{cite}
\usepackage{amsmath,amssymb,amsfonts}
\usepackage{algorithmic}
\usepackage{graphicx}
\usepackage{textcomp}
\usepackage{xcolor}
\usepackage[hyphens]{url}

\usepackage[normalem]{ulem}
\usepackage{fancyhdr}
\usepackage{mathptmx}
\usepackage{algorithm}
\usepackage{xspace}
\usepackage[font=footnotesize,labelfont=bf]{caption}
\usepackage{subfig}
\usepackage{mathtools}
\usepackage{xcolor}
\usepackage{pifont}
\usepackage{color}
\usepackage{soul}
\usepackage{multirow}
\soulregister\cite7
\soulregister\ref7
\soulregister\fig7
\soulregister\todo7
\soulregister\sout7
\soulregister\pageref7
\usepackage{color}
\usepackage{colortbl}
\usepackage{multirow}
\usepackage{makecell}
\usepackage{wrapfig,lipsum}
\usepackage{tikz}
\usetikzlibrary{fit,calc}

\def\BibTeX{{\rm B\kern-.05em{\sc i\kern-.025em b}\kern-.08em
    T\kern-.1667em\lower.7ex\hbox{E}\kern-.125emX}}

\newcommand{\ignore}[1]{}
\newcommand{\todo}[1]{\textcolor{red}{#1}\xspace}

\newcommand{\old}[1]{}
\newcommand{\fig}[1]{Figure~\ref{#1}}
\newcommand{\sect}[1]{Section~\ref{#1}}
\newcommand{\tab}[1]{Table~\ref{#1}}

\newcommand{\eqn}[1]{Equation~\ref{#1}}

\newcommand{\spgemm}[0]{SpDeGEMM\xspace}
\newcommand{\proposed}[0]{GROW\xspace}
\newcommand{\ibufsparse}[0]{I-BUF$_{sparse}$\xspace}
\newcommand{\ibufdense}[0]{I-BUF$_{dense}$\xspace}
\newcommand{\obuf}[0]{O-BUF$_{dense}$\xspace}

\newcommand\blfootnote[1]{%
\begingroup
\renewcommand\thefootnote{}\footnote{#1}%
\addtocounter{footnote}{-1}%
\endgroup
}

\title{\LARGE GROW: A Row-Stationary Sparse-Dense GEMM Accelerator for Memory-Efficient Graph Convolutional Neural Networks} 

\begin{document}

\author{

\IEEEauthorblockN{Ranggi Hwang\IEEEauthorrefmark{2}\IEEEauthorrefmark{4}\hspace{1em} Minhoo Kang\IEEEauthorrefmark{2}\IEEEauthorrefmark{4}\hspace{1em} Jiwon Lee\IEEEauthorrefmark{2}
\hspace{1em} Dongyun Kam\IEEEauthorrefmark{1} \hspace{1em} Youngjoo Lee\IEEEauthorrefmark{1} \hspace{1em} Minsoo Rhu\IEEEauthorrefmark{2}\IEEEauthorrefmark{3}}

\vspace{1em}

\IEEEauthorblockA{\IEEEauthorrefmark{2}KAIST\\School of Electrical Engineering
\\\texttt{\{ranggi.hwang, minhoo.kang, jiwon.lee, mrhu\}@kaist.ac.kr}}
\\
\IEEEauthorblockA{\IEEEauthorrefmark{1}POSTECH\\Department of Electrical Engineering
\\\texttt{\{rkaehddbs, youngjoo.lee\}@postech.ac.kr}}

}

\maketitle
\thispagestyle{empty}
\pagestyle{empty}

%%%%%% -- PAPER CONTENT STARTS-- %%%%%%%%

\input{tex/abstract}

\blfootnote{
\\
\IEEEauthorrefmark{4} Co-first authors who contributed equally to this research.\\
\IEEEauthorrefmark{3} Corresponding author.
}

\input{tex/intro}

\input{tex/background}

\input{tex/related}

\input{tex/motivation}

\input{tex/proposed}

\input{tex/methodology}

\input{tex/evaluation}

\input{tex/discussion}

\input{tex/conclusion}

%%%%%%% -- PAPER CONTENT ENDS -- %%%%%%%%

\section*{Acknowledgment}
This research is partly supported by
the National Research Foundation of Korea (NRF) grant funded by the Korea government(MSIT) (NRF-2021R1A2C2091753),
Institute of Information \& communications Technology Planning \& Evaluation (IITP) grant funded by the Korea government(MSIT) (No. 2022-0-01037, Development of High Performance Processing-in-Memory Technology based on DRAM),
Samsung Advanced Institute of Technology,
and by the Samsung Electronics, Co., Ltd.
We also appreciate the support from the IC Design Education Center (IDEC), Korea, for the EDA tools.
Minsoo Rhu is the corresponding author.

%%%%%%%%% -- BIB STYLE AND FILE -- %%%%%%%%
\bibliographystyle{IEEEtranS}
\bibliography{refs}
%%%%%%%%%%%%%%%%%%%%%%%%%%%%%%%%%%%%

\end{document}

%% file: tex/abstract.tex
\begin{abstract}

Graph convolutional neural networks (GCNs) have emerged as a key technology in
various application domains where the input data is relational.  A unique
property of GCNs is that its two primary execution stages, aggregation and
combination, exhibit drastically different dataflows. Consequently, prior GCN
accelerators tackle this research space by casting the aggregation and
combination stages as a series of sparse-dense matrix multiplication.  However,
						prior work frequently suffers from inefficient data movements,
						leaving significant performance left on the table. We 
						present \proposed, a GCN accelerator based on 
						Gustavson's algorithm to architect a row-wise product based
						sparse-dense GEMM accelerator.
GROW co-designs the software/hardware
						that strikes a balance in locality and parallelism for GCNs,
reducing
the average memory traffic by $2\times$, and achieving an average
$2.8\times$ and $2.3\times$ improvement in performance and 
energy-efficiency, respectively.

\end{abstract}

%% file: tex/intro.tex
\section {Introduction}
\label{sect:intro}

Machine learning (ML) based on deep neural network (DNN) 
primarily target Euclidean data
(e.g., image, audio, and text) which are represented as dense vectors. However,
	an increasing number of application domains where the input data is
	relational (e.g., social networks, knowledge graphs, user-item graphs) 
	are 
	gaining significant traction, for which graphs are more powerful
representations. As such, several studies explored the possibility of
	applying DNNs to graph
	representations~\cite{gcn,graphsage,gat}.  Graph convolutional neural
	networks (GCNs)~\cite{gcn,hygcn,awbgcn,gcnax} are one of those prominent
	approaches which extend the concept of convolutions for graph data, achieving
	state-of-the-art performance in areas of e-commerce, social network analysis,
	molecular graphs, 	advertisement, 
	etc~\cite{yang2019aligraph,lerer2019pytorch,de2018molgan}. 
	
	With the rapid development of GCNs, designing domain-specific
	architectures for GCNs have received significant
	attention~\cite{hygcn,awbgcn,gcnax}. The graph convolutional layers typically
	take up the majority of execution time in GCN inference through two primary
	stages: \emph{Aggregation} and \emph{Combination}. The dataflow of
	combination phase resembles that of conventional DNN algorithms, exhibiting
	dense compute and highly regular memory accesses. 
In contrast, the aggregation phase exhibits
	the typical graph processing's characteristics, showing highly sparse and
	irregular memory access patterns. Such conflicting and heterogeneous
	requirement of GCN dataflow imposes a unique challenge for GCN accelerators,
	as both the irregularity of aggregation and the regular dataflow of
	combination must simultaneously be exploited for high efficiency.

	Given this landscape, recent literature proposed dedicated GCN accelerators
	that achieve substantial energy-efficiency improvements than general-purpose
	GPUs or TPUs. The pioneering work on HyGCN~\cite{hygcn} is one of the first
	domain-specific accelerator for GCN, employing two separate accelerator
	engines for aggregation and combination, respectively. While effective, HyGCN
	can suffer from under-utilization because of the load-imbalance between the
	two engines. As such, two recent GCN accelerators such as AWB-GCN~\cite{awbgcn}
	and GCNAX~\cite{gcnax} employ a \emph{unified} compute engine that can handle
	both the aggregation and combination stages. The key intuition behind these
	two works is that the aggregation and combination stages of graph convolutional layer can be permuted into two
	consecutive sparse-dense GEMM (\spgemm) operations by changing the execution
	order of its matrix multiplication operations (\sect{sect:gcn_as_spgemm}). This allows a single,
	unified microarchitecture design, tailored for \spgemm, to execute both the
	aggregation and combination phases, successfully addressing HyGCN's
	load-imbalance issue.  As we explore in this paper, however, prior
	unified \spgemm based GCN accelerators fall short 
	because they are not able to exploit the unique
	sparsity patterns of aggregation and combination, resulting in significant
	waste in memory bandwidth utilization.  A {\bf key observation} of our study
	is that the
	input sparse matrices of graph convolutional layer's two \spgemm 
	exhibit drastically different levels of sparsity, i.e., a \emph{heterogeneous}
	mix of sparsity where the sparse matrix of aggregation is typically orders of
	magnitude sparser than that of combination (\sect{sect:sparsity_level}), presenting
	unique opportunities to reduce memory traffic and data movements. Existing
	GCN accelerators for \spgemm, however, fail to reap out such opportunity as they
	employ a rigid computational dataflow in handling both of the two \spgemm,
	resulting in an average $73\%$ of memory bandwidth waste
	when evaluated with state-of-the-art GCNAX design (\sect{sect:gcnax_bottleneck}).

	To this end, this paper presents our {\bf G}CN accelerator based on a {\bf
		Row}-stationary \spgemm dataflow (\proposed). To unlock the full
		potential of GCNs, \proposed is designed with the following key
		features. 

		\begin{enumerate}

\item Similar to AWB-GCN/GCNAX, \proposed is based on a unified
			microarchitecture tailored for \spgemm, 
			enabling seamless execution of both aggregation combination
			phases. Unlike prior \spgemm accelerators, however, we co-design the software/hardware
			architecture (as discussed below) to minimize its data
			movements, significantly enhancing the performance of the memory-bound
			\spgemm.

		\item \proposed employs a \emph{row-stationary} dataflow based on the
		\emph{row-wise product} matrix multiplication (aka Gustavson's
				algorithm~\cite{gustavson}), allowing both flexible and fine-grained
		adaptation to the heterogeneous sparsity patterns of aggregation and
		combination. Compared to GCNAX, \proposed's row-stationary dataflow
		drastically reduces the memory bandwidth waste, especially during the
		aggregation phase which dominates GCN's inference time.

		\item While \proposed's row-stationary dataflow helps better adapt to the
		heterogeneous sparsity patterns of \spgemm's sparse matrices, it does
		incurs more irregular reuse of the dense matrices. \proposed employs a
		\emph{graph partitioning algorithm}, a software-level preprocessing technique targeting the
		adjacency matrix of GCN (i.e., the sparse matrix in aggregation), which
		significantly improves the locality of row-stationary dataflow.

		\item Coupled with \proposed's graph partitioning algorithm, we propose a
		\emph{multi-row stationary runahead execution} model, a hardware
		microarchitecture co-designed with the graph partitioning scheme as means
		to maximize memory-level parallelism and overall throughput.

		\end{enumerate}

We implement \proposed and evaluate its performance and energy-efficiency
across a wide range of GCNs. Compared to state-of-the-art GCNAX, \proposed
reduces the average memory traffic by $2\times$, and achieves an average
$2.8\times$ and $2.3\times$ improvement in performance and 
energy-efficiency, respectively.

%% file: tex/background.tex
\section{Background}
\label{sect:background}

\subsection{Fundamentals of GCN}
 \label{sect:gcn_intro}

\begin{figure}[t!] \centering
\includegraphics[width=0.485\textwidth]{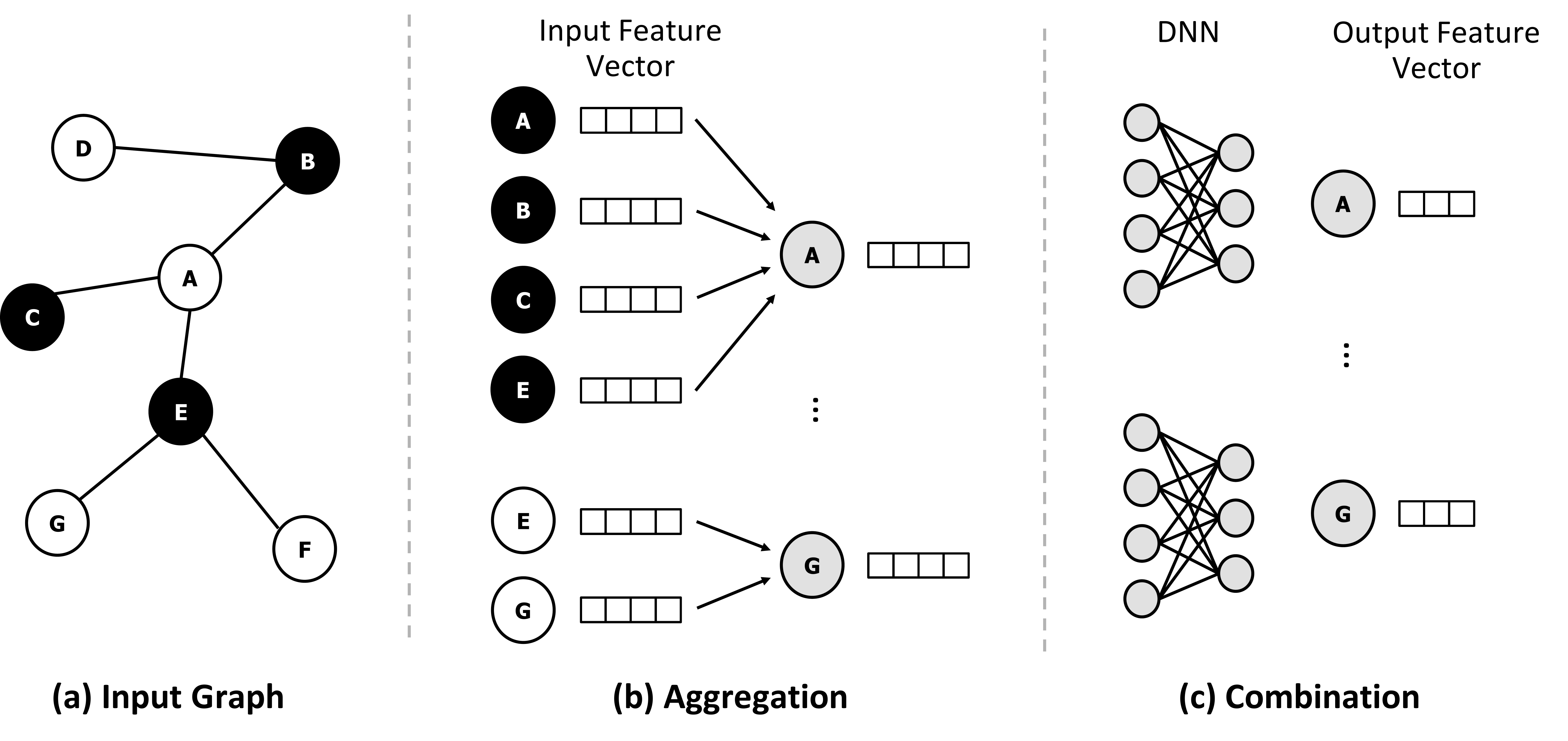}

\caption{
The aggregation and combination stages of GCN.
}
\vspace{-1.25em}
\label{fig:background_gcn}
\end{figure}

Graph neural networks (GNNs) 
can extract meaningful features by learning
both the representation of each objects (i.e., graph nodes) as well as the relationship 
across different objects (i.e., the edges that connect nodes). GCNs 
apply the concept of convolutions for the graph's 
feature extraction. GCNs adopt a neighborhood aggregation scheme, where the output feature
vector of each vertex is derived by recursively aggregating and transforming 
the input feature vectors
of its neighbor vertices. In \fig{fig:background_gcn}, we show the two key execution
phases of GCN inference: \emph{Aggregation} and \emph{Combination}. After \emph{N}
iterations of aggregation and combination, the target vertex is represented by its
final output feature vector, which encapsulates the unique structural information of
the vertex's \emph{N}-hop neighborhood. \eqn{eqn:gcn_inf} shows the layer-wise
forward propagation of a single graph convolutional layer:

\begin{equation}
\label{eqn:gcn_inf}
X^{(l+1)}=\sigma(AX^{(l)}W^{(l)}) 
\end{equation}

 $A$ denotes the adjacency matrix of the graph where each row represents the
 connection of a vertex with all the other vertices within the graph. $X^{(l)}$
 is a matrix containing all the input feature vectors of all the vertices in
 layer-$l$, i.e., each column of $X$ represents a feature while each row
 denotes a given vertex's feature vector. 
$W$ contains the GNN's model parameters that are subject for \emph{training}.
	Training models for small to medium sized graphs (where the entire
		graph and intermediate states of all graph nodes can be stored in memory)
	involve operating
on the full graph Laplacian, 
whereas large-scale GNNs that do not fit within memory
typically involve
a graph node ``sampling'' process~\cite{graphsage}.
Depending on what the downstream task
the GNN is being trained for (e.g., node/edge prediction~\cite{david_nips, alex_nips, gcn},
graph clustering~\cite{rex_nips},
recommendation	models~\cite{hanjun_icml}),
		the input queries that constitute the training as well as test
dataset can vary significantly (e.g., node IDs for node prediction, user/item IDs
		for recommendation models in E-commerce).
 $\sigma$() denotes the non-linear
 activation function such as ReLU. Because the nodes with more neighbors tend
 to have larger values under feature extraction, $A$ is typically normalized to
 prevent it from changing its scale. The 
 normalization of $A$, however, can be done offline as a one-time preprocessing stage, 
 so the remainder of this paper
 assumes $A$ to denote the normalized version of it. 

\begin{table*}[t]
    \centering
    \caption{Structure and key features of the graph datasets and GCN configuration. Graphs are sorted based on the number of nodes (from left to right). In this paper, we explore an even mix of both small-scale (Corea/Citeseer/Pubmed/Flickr) and large-scale datasets (Reddit/Yelp/Pokec/Amazon) to highlight different effects \proposed have as the scale of graph dataset is changed.}
\scriptsize
    \begin{tabular}{|c||c|c|c|c|c|c|c|c|}
    \hline
    \textbf{Datasets} & \textbf{Cora} & \textbf{Citesee}r & \textbf{Pubmed} & \textbf{Flickr} & \textbf{Reddit} & \textbf{Yelp} & \textbf{Pokec} & \textbf{Amazon}\\
    \hline
    \hline
    \# of Nodes & 2,708 & 3,327 & 19,717 & 89,250 & 232,965 & 716,847 & 1,632,803 & 2,449,029 \\
    \hline
    \# of Edges & 13,264 & 12,431 & 108,365 & 989,006 & 114,848,857 & 13,954,819 & 46,236,731 & 126,167,309 \\
    \hline
    Average degree & 4.90 & 3.74 & 5.50 & 11.1 & 493 & 19.5 & 28.3 & 51.5 \\
    \hline
    Feature length & 1,433-16-7 & 3,703-16-6 & 500-16-3 & 500-64-7 & 602-64-41 & 300-64-100 & 60-64-48 & 100-64-47 \\
    \hline
    Density of A & $1.81\times10^{-3}$ & $1.12\times10^{-3}$ & $2.79\times10^{-4}$ & $1.24\times10^{-4}$ & $2.12\times10^{-3}$ & $2.72\times10^{-5}$ & $1.73\times10^{-5}$ & $2.10\times10^{-5}$\\
    \hline
    Density of $X^{(0)}$ & 1.27\% & 0.85\% & 10.0\% & 46.4\% & 100\% & 100\% & 39.9\% & 99.0\%\\
    \hline
    Density of $X^{(1)}$ & 78.0\% & 89.1\% & 77.6\% & 77.2\% & 63.9\% & 77.2\% & 77.2\% & 77.2\%\\
    \hline
    Density of W & 100\% & 100\% & 100\% & 100\% & 100\% & 100\% & 100\% & 100\% \\

    \hline
    \end{tabular}
    \label{tab:model_config}
		\vspace{-1.5em}
\end{table*}

\subsection{GCN as a Sparse-Dense GEMM Operation}
\label{sect:gcn_as_spgemm}

\begin{figure}[t!] \centering
\includegraphics[width=0.485\textwidth]{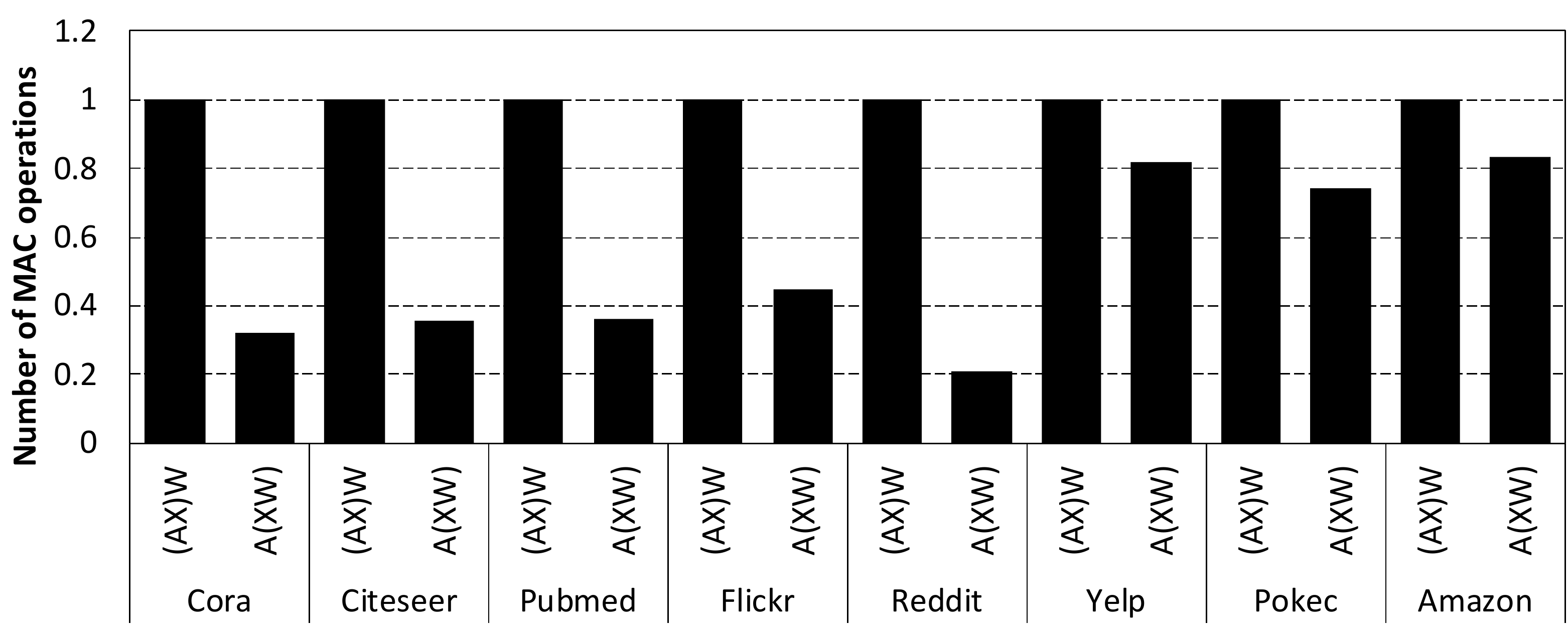}
\caption{
Normalized number of MAC operations depending on the execution order of two GEMMs, ($A$$\times$$X$)$\times$$W$ vs. $A$$\times$($X$$\times$$W$).
}
\vspace{-1em}
\label{fig:background_gemm_order}
\end{figure}

A key computation pattern of a GCN layer is the two-stage
matrix-multiplication, $A$$\times$$X$$\times$$W$. There are two possible
execution orders: ($A$$\times$$X$)$\times$$W$ and $A$$\times$($X$$\times$$W$).
Prior work
on AWB-GCN~\cite{awbgcn} observed that the order in which $A$$\times$$X$$\times$$W$
is conducted has significant impact on the total number of MAC operations conducted. As shown
in \tab{tab:model_config}, $A$ tends to be extremely large and sparse,
	 $X$ being modestly sized and can be somewhat sparse but also be dense depending on the workload, and $W$ being small and highly dense.
	 Consequently, adopting the ($A$$\times$$X$)$\times$$W$ execution order involves
	 multiplying the sparse $A$ and sparse-or-dense $X$ first, producing a largely sized
	 dense matrix, which is subsequently multiplied by another dense matrix $W$, leading to high
	 computations. The alternative $A$$\times$($X$$\times$$W$) on the other hand
	 can be handled as two consecutive sparse-dense matrix multiplications (\spgemm) which can 
	 lead to smaller amount of computations (\fig{fig:background_gemm_order})~\cite{hygcn,gcnax}. Another important
	 advantage of the aforementioned $A$$\times$($X$$\times$$W$) execution order is
	 that the GCN layer can be handled with a single sparse-dense
	 GEMM operation, making it possible to execute the two-stage \spgemm end-to-end over
	 a \emph{unified} \spgemm accelerator design.

\subsection{State-of-the-art Accelerators for GCN}
\label{sect:related_GCNs}

HyGCN~\cite{hygcn} is one of the first GCN accelerators that address the hybrid
dataflow of both aggregation and combination. Specifically, HyGCN employs two
separate engines, an aggregation engine optimized for the
sparse-sparse GEMM of ($A$$\times$$X$) and a combination engine handling the
subsequent dense-dense GEMM of ($AX$$\times$$W$). While HyGCN provides
significant speedups vs. CPUs/GPUs, it falls short on two
important aspects.  First, one of the two  engines in HyGCN can suffer from
under-utilization due to load-imbalance. Second, as discussed in
\sect{sect:gcn_as_spgemm}, the execution order of ($A$$\times$$X$)$\times$$W$
can lead to a suboptimal dataflow with much higher computation demands than
$A$$\times$($X$$\times$$W$), leaving significant performance left on the table. 

Consequently, AWB-GCN~\cite{awbgcn}
employs the $A$$\times$($X$$\times$$W$) execution order of
graph convolution, presenting a unified accelerator microarchitecture design.
By permuting the two consecutive matrix multiplication operations into a \spgemm
operator, AWB-GCN achieves significant reduction in computational requirements
with high speedups compared to HyGCN. More recently, GCNAX~\cite{gcnax} introduces
further dataflow optimizations on top of AWB-GCN using reconfigurable loop ordering,
				loop fusion, and efficient tiling strategies, achieving significant energy-efficiency
				improvements against both HyGCN and AWB-GCN. 
Given its robust accelerator design and  state-of-the-art performance, the remainder
of this paper assumes
the $A$$\times$($X$$\times$$W$) execution order for GCN inference, using
GCNAX as the baseline \spgemm based GCN accelerator. Additionally, we henceforth refer to the
$A$$\times$($XW$) as the aggregation phase and $X$$\times$$W$ as the combination
phase for clarity of explanation.

%% file: tex/related.tex
\section{Related Work}
\label{sect:related}

\begin{table*}[t!]
	\scriptsize
	  \centering
	  \caption{Comparison between GROW and related work.}
	\vspace{-0em}
	  \begin{tabular}{|l|c|c|c|c|c|}
		\hline & GCNAX~\cite{gcnax} &  AWB-GCN~\cite{awbgcn} & HyGCN~\cite{hygcn} & GAMMA~\cite{gamma}& Ours \\
		\hline
			{\bf Compute engine} & Unified & Unified & Heterogeneous & Unified& Unified\\
		\hline
			{\bf Operend} & Sp-De & Sp-De & Sp-De / De-De & Sp-Sp & Sp-De\\
		\hline
			{\bf Product type} & Outer product & Inner product & \makecell{Outer product /\\Inner product}   & Row-wise product & Row-wise product\\
		\hline
			{\bf Compress format} & CSC & CSR & CSC & CSR & CSR\\
		\hline
			{\bf Preprocessing scheme} & None & None & \makecell{Graph partitioning /\\ Vertex interval \& edge sharding }&
			\makecell{Row Reordering \\ \& Coordinate-space tiling}& \makecell{HDN caching \& graph partitioning}\\ 
		\hline
	  \end{tabular}
	\vspace{-0.8em}
	  \label{tab:comparison}
	\end{table*}
	
There is a large body of prior work that seek acceleration of 
sparse DNNs~\cite{gnn:survey,auten2020hardware,grip,graphact,graphsaint,zhang2020hardware,centaur:hwang,engn,cambricong,tensaurus,awbgcn,gcnax,hygcn,boostgcn,greta,huang:sc20:gespmm,gnnadvisor,smartsage:isca2022,scratchpipe:isca2022, trim:micro, tcasting, trim:cal, tensordimm, rhu:2018:cdma, scnn}.
	We summarize a subset of other related work in \tab{tab:comparison}
that explores 1) alternative measures for
accelerating GCNs, 2) locality-enhancing mechanisms for graph
analytics, and 3) accelerator for sparse linear algebra.

{\bf Accelerating GCNs using ASIC/FPGA/GPU.}
Along with HyGCN~\cite{hygcn}, EnGN~\cite{engn} is one of those first accelerators targeting
GCNs, employing an output stationary dataflow with sequential pipelining to
interleave the computation of aggregation and combination for high throughput.
Cambricon-G~\cite{cambricong} is another recent GCN accelerator employing a 
multi-dimensional cuboid engine. 
GRIP~\cite{grip} is an accelerator based on the GReTA~\cite{greta} programming
abstraction whose architecture design and dataflow is similar to HyGCN.
GraphACT~\cite{graphact} is a CPU-FPGA based accelerator targeting GCN training
and GNNAdvisor~\cite{gnnadvisor} proposes an efficient software runtime system
for GNN acceleration in GPUs. In general, the contribution of \proposed is orthogonal
to these prior work.

{\bf Techniques to enhance locality in graph analytics.} Graph analytics are
well-known for its irregular and sparse memory accesses exhibiting low
locality. Consequently, several locality-enhancing solutions have been
proposed in past literature for graph analytics
(e.g., intelligent graph partitioning~\cite{metis,graclus}, vertex
		reordering~\cite{zhang2018degree}). For instance, Zhang et
	al.~\cite{zhang2018degree} proposes a node degree-aware vertex reordering
	mechanism that helps improve locality for graph traversal algorithms.
	\proposed builds upon these prior work to address the irregular reuse of
	dense matrices in GCN's \spgemm based inference, which we detail in
	\sect{sect:clustering}.

{\bf Accelerators for sparse linear algebra.} Several recent studies explored
the design of domain-specific architectures for sparse linear algebra.
OuterSPACE~\cite{outerspace} and SpArch~\cite{sparch} are two recent works that
propose hardware accelerators for sparse-sparse GEMMs using an outer-product
approach.  ExTensor~\cite{extensor} is another recent study that applies the
inner-product approach for sparse GEMMs.  Tensaurus~\cite{tensaurus} is a
versatile accelerator targeting both dense and sparse tensor factorizations.
Most recently, MatRaptor~\cite{matraptor} and GAMMA~\cite{gamma} 
explored the
efficacy of utilizing Gustavson's algorithm for accelerating sparse-sparse
GEMMs. Similar to GROW, these two works employ a row-wise product based
dataflow for minimizing data movements in this memory bandwidth limited
algorithm, achieving significant energy-efficiency
improvements. However, MatRaptor and GAMMA target generic sparse-sparse GEMM
algorithms so it is not optimized to leverage the unique algorithmic properties
of GCNs (e.g., the power-law distribution of graph datasets).  Furthermore, the
GCN we study is formulated as a sparse-dense GEMM, unlike the sparse-sparse
GEMM these prior work focuses on. In \sect{sect:eval_gamma}, we quantitatively
demonstrate GROW's merits over both MatRaptor and GAMMA.  Overall, the key
contributions of GROW are orthogonal to these prior studies.

%% file: tex/motivation.tex
\section{Motivation} 
\label{sect:motivation}

\sethlcolor{green}
\begin{figure}[t!] 
\centering
\subfloat[]{
\includegraphics[width=0.485\textwidth]{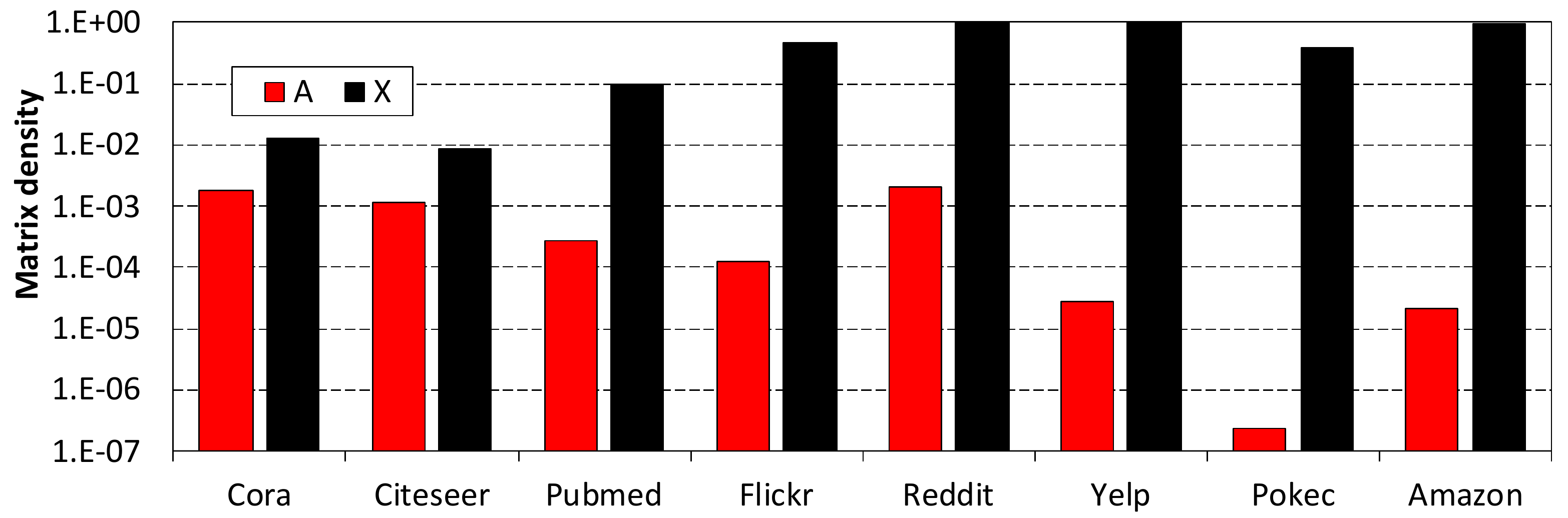}
	\label{fig:}
}
\vspace{0.5em}
\subfloat[]{
	\includegraphics[width=0.485\textwidth]{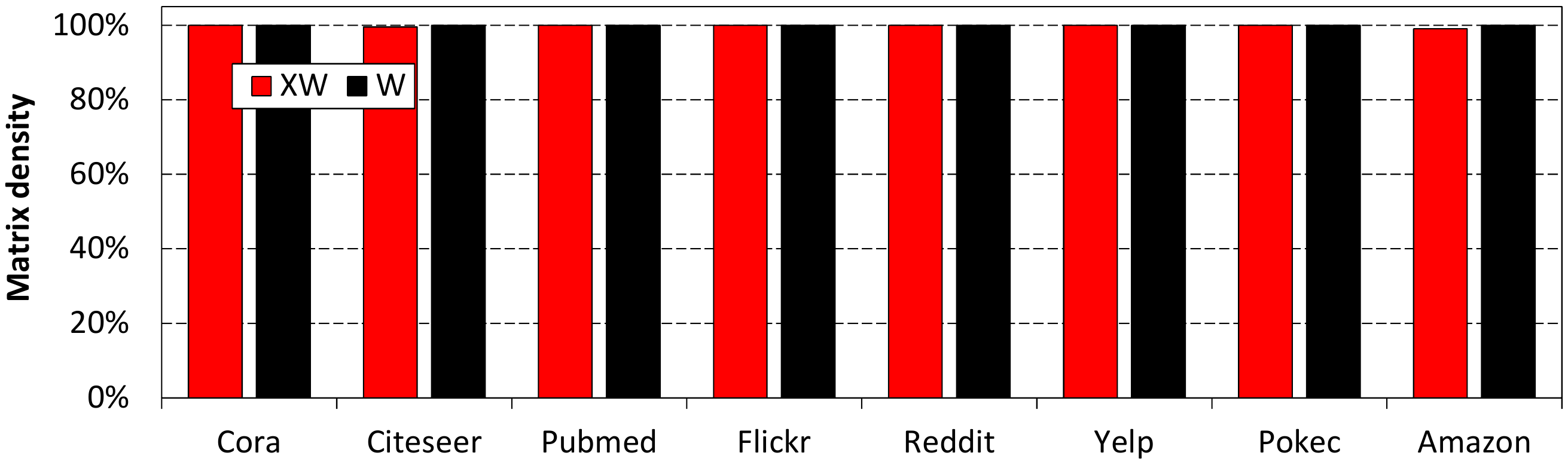}
	\vspace{-1.5em}
	\label{fig:}
}
\caption{ 
Density of (a) sparse matrices ($A$, $X$) and (b) dense matrices ($XW$, $W$) in the aggregation and combination stages.
}   
\vspace{-1.25em}
\label{fig:sparse_matrix_density}
\end{figure}

\subsection{Heterogeneous Levels of Sparsity in Graph Datasets}
\label{sect:sparsity_level}

A key limitation of prior \spgemm accelerators is that they do not exploit the
different levels of sparsity manifested during aggregation and combination,
					simply treating them as a generic \spgemm.
					\fig{fig:sparse_matrix_density} illustrates the density of input
					matrices during aggregation and combination. First thing to note is
					that the right-hand side matrices (i.e., $XW$ in $A$$\times$$XW$ and $W$ in
							$X$$\times$$W$)  of aggregation and
					combination are all extremely dense across all the graph datasets.
					However, the left-hand side matrices of the two \spgemm operations (i.e., $A$ in $A$$\times$$XW$ and $X$ in
							$X$$\times$$W$) exhibit drastically different sparsity levels.
					That is, the sparse matrix in aggregation ($A$) is universally sparse
					with orders of magnitude higher sparsity than that of combination
					($X$). 
					The	sparsity of matrix $X$, however, is mixed and oftentimes non-existent to
					begin with, i.e., highly dense for Reddit, Yelp, and Amazon. 

Consequently, while the aggregation phase of GCN
				(i.e., $A$$\times$$XW$) can easily be generalized as a sparse-dense matrix multiplication, the combination phase can either be sparse-dense or dense-dense depending on 
				the nature of the input feature vector, i.e., matrix $X$.
					Despite such heterogeneous
					and mixed levels of sparsity, however, the \spgemm accelerator in
					GCNAX (as well as AWB-GCN) applies a rigid computational dataflow
					strictly tailored for the sparse-dense matrix multiplication for both aggregation
					and combination
					stages.  
					We show that failing to incorporate such heterogeneous
					sparsity results  in needless overprovisioning of on-chip
					buffers as well as significant waste in memory bandwidth.

\begin{figure}[t!] 
\centering
\includegraphics[width=0.495\textwidth]{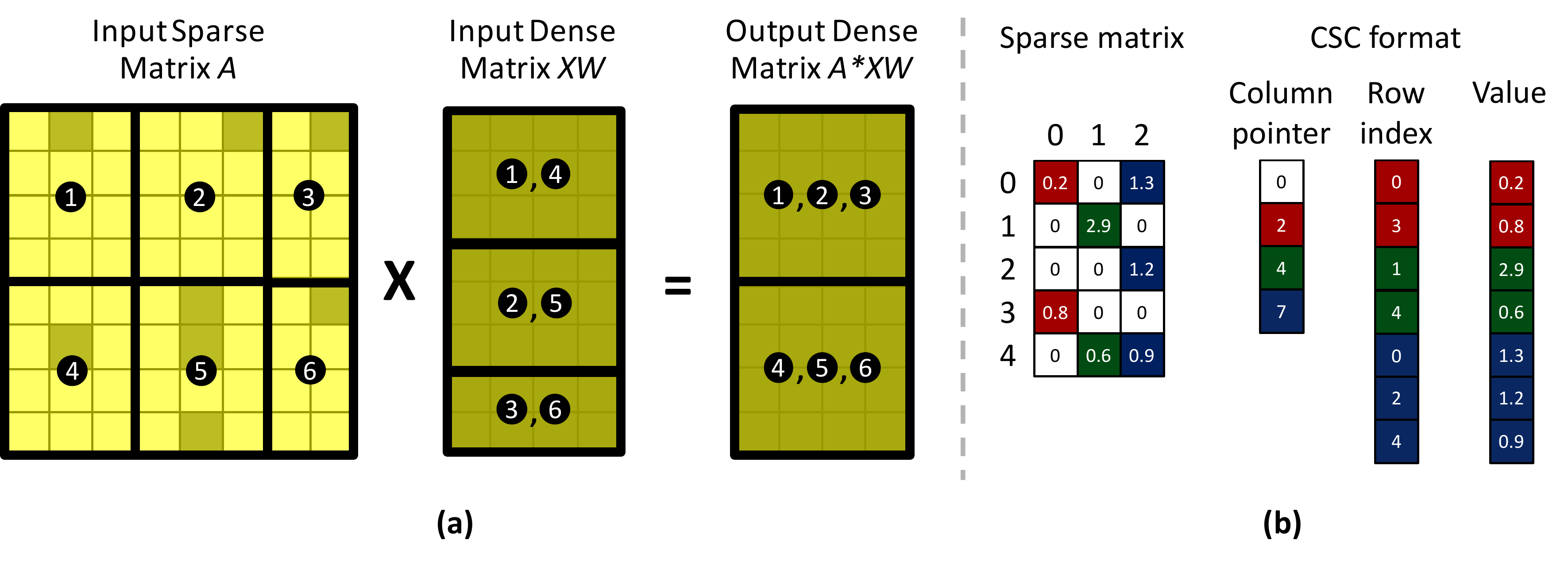}
\caption{ 
(a) GCNAX's matrix tiling strategy for aggregation. The black numbered circles designate the order in which
	the input tiles are fetched (for $A$ and $XW$) and utilized for output tile derivation (for $A$$\times$$XW$). (b) The  input sparse matrix is compressed in a CSC (compressed sparse column) format.
}   
\vspace{-1em}
\label{fig:gcnax_tiling}
\end{figure}

\subsection{Tiling and Its Effect on On-/Off-chip Data Movement}
\label{sect:gcnax_bottleneck}

Tiling matrices to temporarily store a subset of the active working set is a
well-known, proven optimization strategy to reduce off-chip bandwidth
requirement. Improving upon AWB-GCN's tiling scheme, GCNAX proposes a
dynamically reconfigurable loop unrolling and tiling mechanism, the high-level
overview of which is provided in \fig{fig:gcnax_tiling}. By carefully examining
the two \spgemm matrix shapes and graph dataset, GCNAX derives an optimal tile
size for the sparse and dense matrices (tiled $A$ and tiled $XW$ in
		\fig{fig:gcnax_tiling}(a)), which is utilized to derive the output
tile based on an \emph{outer-product} GEMM among the tile $A$ and tile $XW$. Note that the tiled sparse matrix
is compressed in a CSC format (\fig{fig:gcnax_tiling}(b)), which the GCNAX's
memory fetch unit utilizes to only fetch the non-zero elements within the
currently working tile.

Our characterization reveals two important insights. First,
		GCNAX must provision the on-chip buffer size to be large
		enough to sufficiently capture the worst-case working set
		of sparse matrices $A$ and $X$.  As discussed in
		\fig{fig:sparse_matrix_density}(a), however, matrix $X$ in
		several graph datasets are completely dense, so the size
		of the on-chip buffer  must be (over)provisioned to match
		the exact size of GCNAX's chosen tile size \emph{in the
			worst case} that the input tile is completely dense.
			Because the density of matrix $A$ is extremely
			sparse, however, such overprovisioning of on-chip
			buffers are an extreme overkill during aggregation,
			wasting precious on-chip SRAM storage.

\begin{figure}[t!] 
\centering
\subfloat[]{
\includegraphics[width=0.485\textwidth]{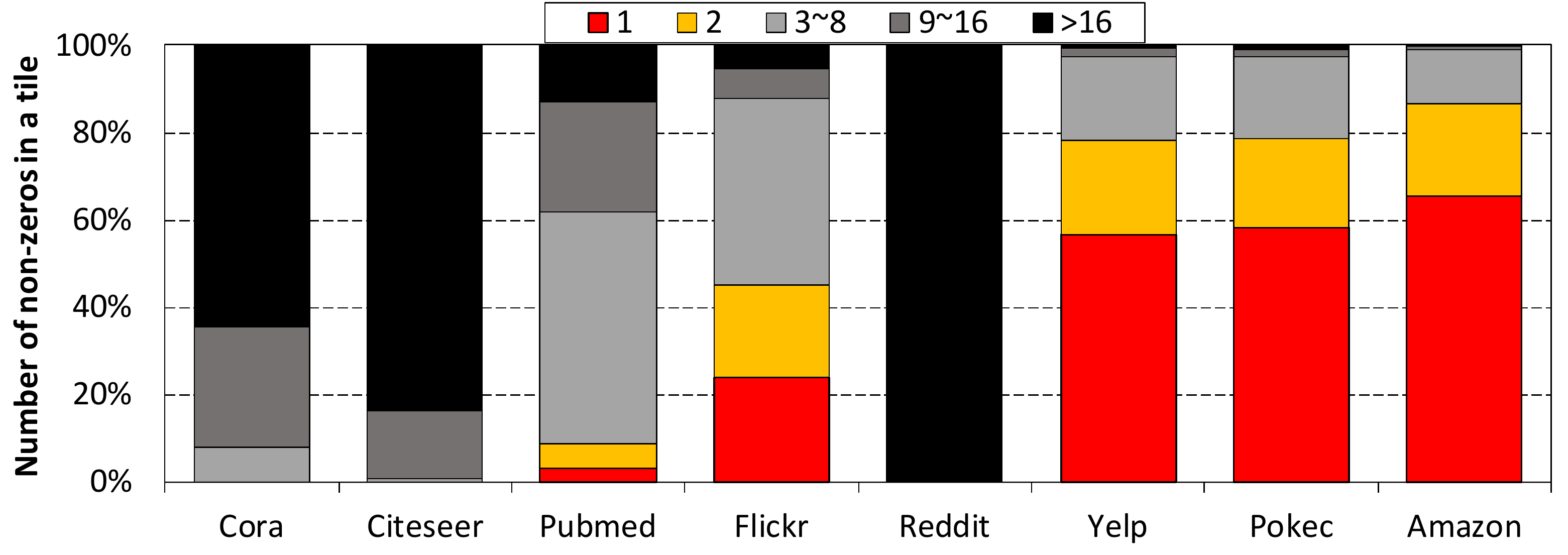}
\vspace{-1.25em}
	\label{fig:}
}
\vspace{0.5em}
\subfloat[]{
	\includegraphics[width=0.485\textwidth]{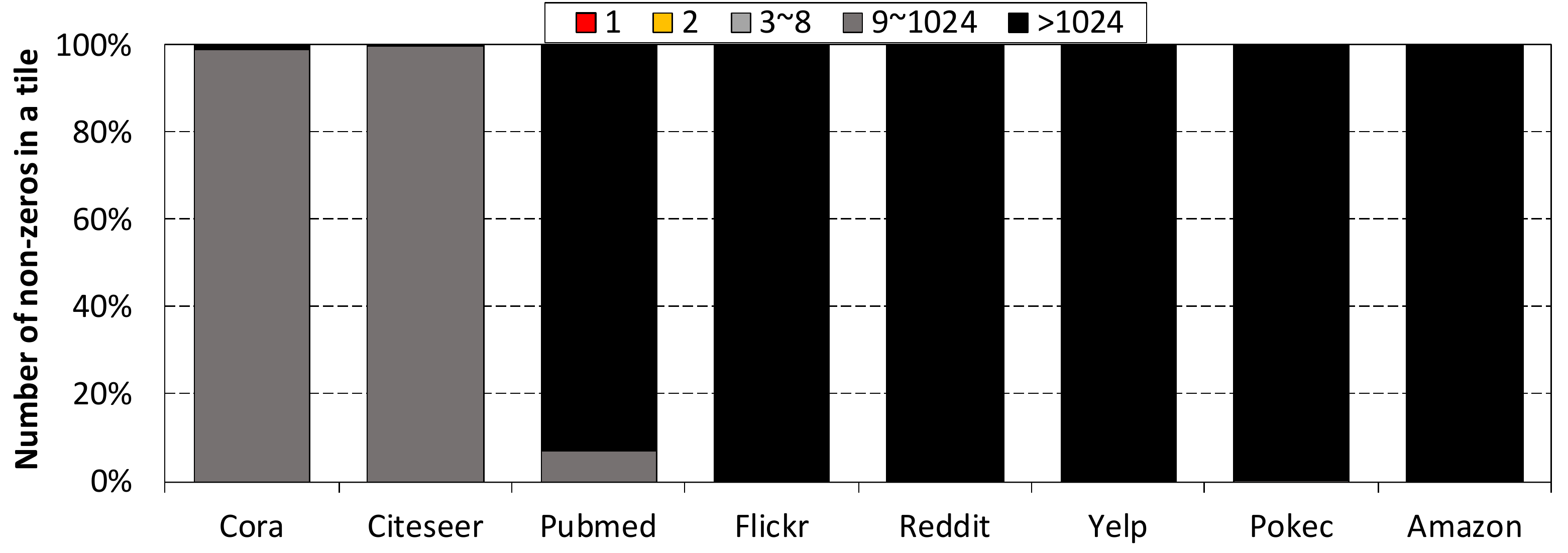}
	\vspace{-1.25em}
	\label{fig:}
}
\caption{ 
The number of non-zero elements within a tile in (a) matrix $A$ (aggregation) and (b) $X$ (combination).
}   
\vspace{-1.5em}
\label{fig:num_nz_in_tile}
\end{figure}

\begin{figure}[t!] \centering
\includegraphics[width=0.485\textwidth]{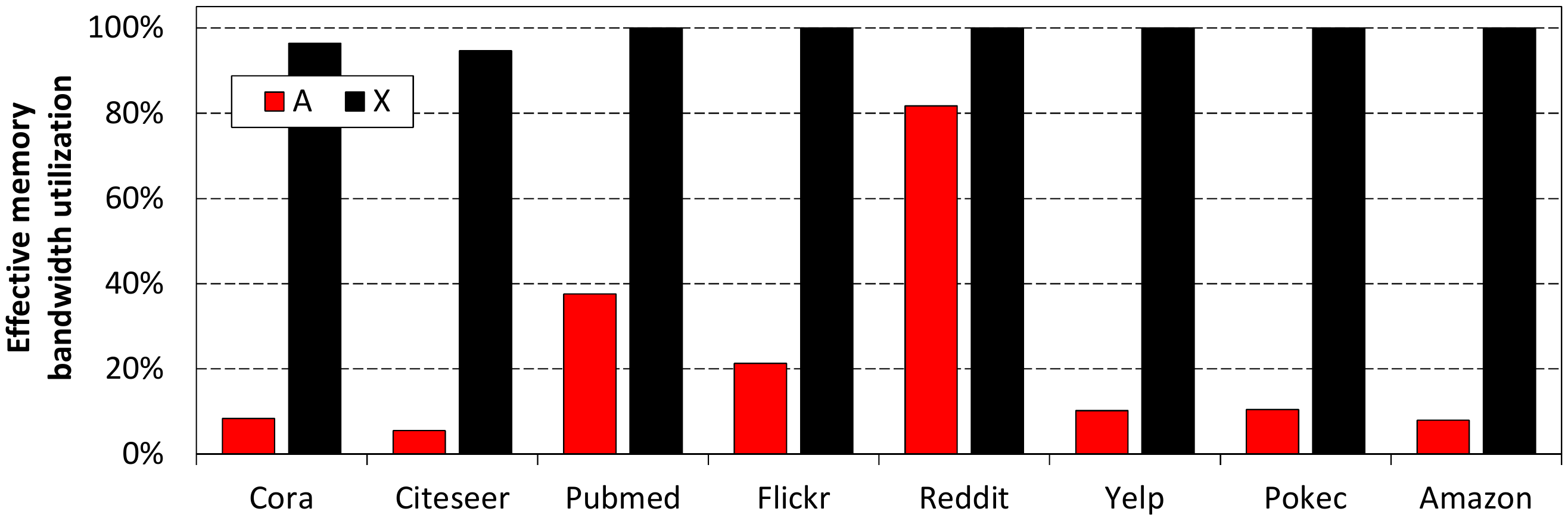}
\caption{
Effective memory bandwidth utilization when fetching the sparse matrices $A$ and $X$. Effective
	memory bandwidth utility is measured by counting how many bytes within the data read from DRAM are fetched {on-demand} assuming a $64$ byte minimum access granularity memory system.
}
\vspace{-0em}
\label{fig:effective_bw}
\end{figure}

\begin{figure}[t!] \centering
\includegraphics[width=0.485\textwidth]{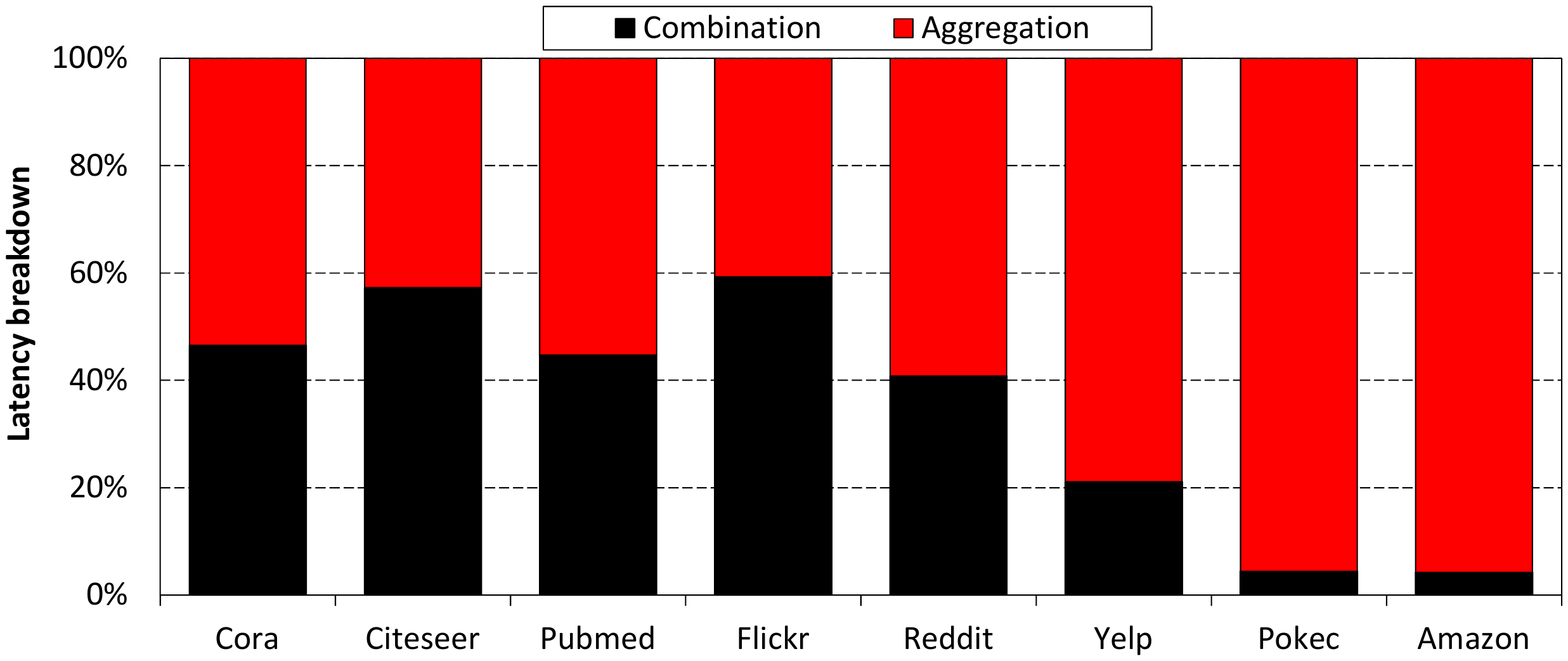}
\caption{
Breakdown of GCNAX's end-to-end inference latency.
}
\vspace{-1.5em}
\label{fig:motivation_gcnax_latency_breakdown}
\end{figure}

Another important observation we make is that the number of non-zero elements
manifested within a tile, for aggregation and combination, exhibits drastically
different characteristics. \fig{fig:num_nz_in_tile} summarizes the number of
non-zero elements existent within the sparse tile fetched for GCNAX's
outer-product \spgemm.  During the combination stage, the number of non-zero
elements within the tiled matrix $X$ are reasonably high, typically much more
than $1024$ non-zeros per tile (\fig{fig:num_nz_in_tile}(b)). Such high
density of non-zero elements enables a dense packing of \emph{effectual} data
within the compressed-sparse CSC format per each tile, achieving high memory
bandwidth utilization when fetching the non-zero elements  from the DRAM
subsystem	(the black bars in \fig{fig:effective_bw}).  Unfortunately, the
sparsity manifested during the aggregation stage exhibits a completely
different pattern.
As depicted in \fig{fig:num_nz_in_tile}(a), the adjacency
matrix $A$ typically contains a very small number
of non-zero elements (e.g., for Yelp/Pokec/Amazon, less than $3$ non-zero elements for more
than $80\%$ of the tiles fetched from DRAM). Consequently, GCNAX's
effective memory bandwidth utilized for fetching effectual non-zero elements
during aggregation becomes significantly low, averaging at $23\%$ (worst
		case $<$$6\%$) bandwidth utilization (the red bars in
			\fig{fig:effective_bw}).  This is because the effectual non-zero elements
		within any given tile of matrix $A$ is typically much smaller than the
		minimum data access granularity of the DRAM subsystem (i.e., $64$ bytes),
		leading to a significant overfetch of useless data for the tiled \spgemm.
		Given the memory bandwidth limited nature of sparse matrix multiplication
		operations~\cite{outerspace,sparch,extensor}, such significant waste in memory throughput
		renders the dataflow of GCNAX suboptimal, leaving significant performance
		left on the table. Overall, the performance of GCNAX gets completely
		bottlenecked by the low memory bandwidth utilization during the aggregation
		stage, spending significantly higher latency during aggregation than during the
		combination stage for large-scale graph datasets (\fig{fig:motivation_gcnax_latency_breakdown}).

%% file: tex/proposed.tex
\section{GROW Architecture and Design} 
\label{sect:proposed}

\subsection{Architecture Overview}
\label{sect:arch_overview}

\begin{figure}[t!] \centering
\includegraphics[width=0.45\textwidth]{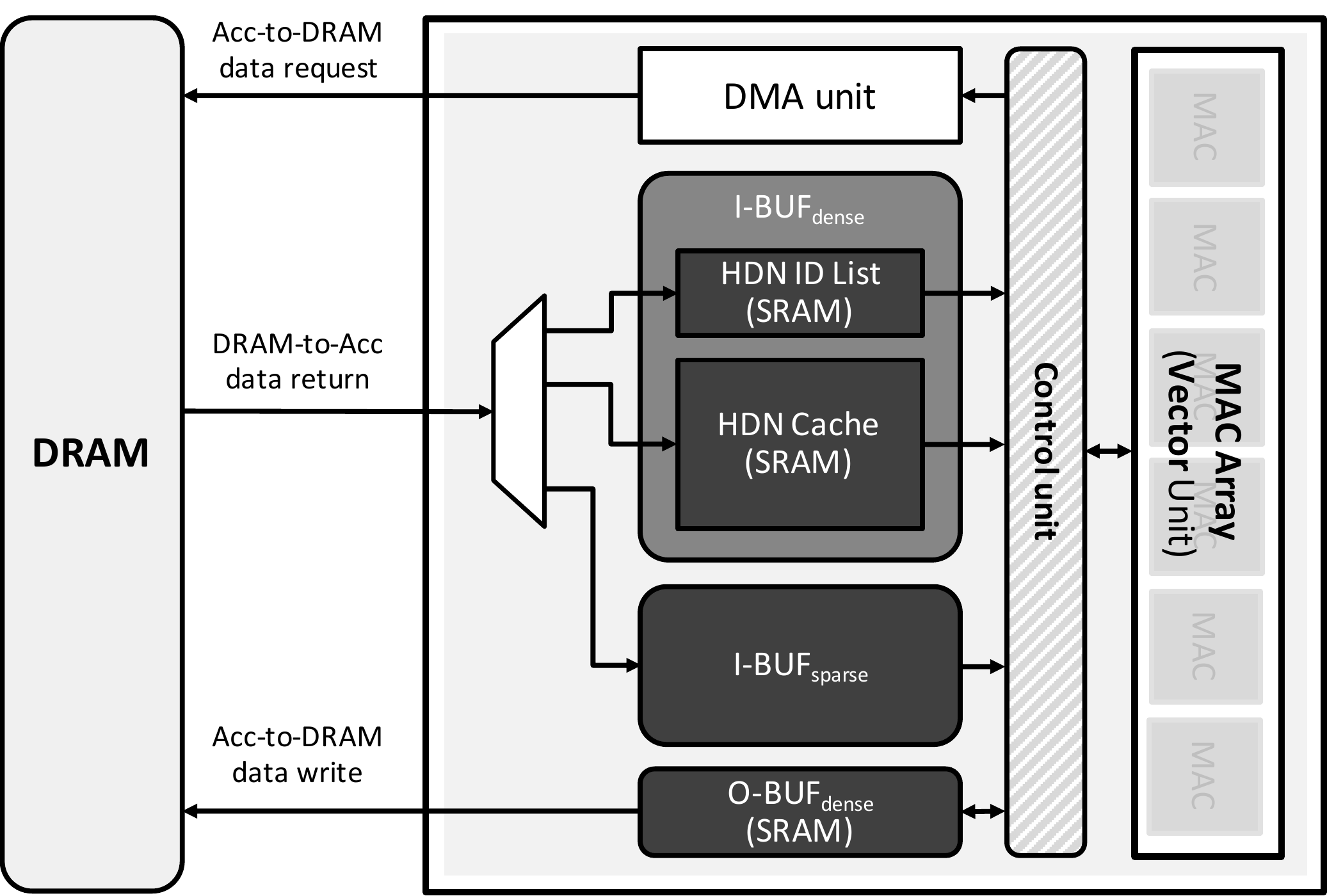}
\caption{
High-level overview of \proposed.
}
\vspace{-0.5em}
\label{fig:proposed_arch}
\end{figure}

\proposed consists of a compute engine based on a vector processor, three sets
of on-chip SRAM (\ibufsparse, \ibufdense, and \obuf) to maximize
locality, a DMA unit that orchestrates on-/off-chip data movements, and the
main control unit (\fig{fig:proposed_arch}).  The control unit populates the
\ibufsparse and \ibufdense buffers with the sparse ($A$ and $X$) and dense
($XW$ and $W$) matrices of aggregation and combination stages and conducts
the row-wise product based GEMM operation using vector processing, the output
of which is stored into the \obuf in row granularity.  The \ibufdense is
functionally split into two subblocks, a high-degree node (HDN) \emph{cache}
that captures high locality dense matrix rows and a CAM (content addressable
		memory) based \emph{buffer} that stores the list of top-$N$ high-degree
node's IDs.  The sparse matrices $A$ and $X$ are compressed in CSR format
while matrices $XW$ and $W$ are stored without compression in a dense fashion.

\subsection{Dataflow}
\label{sect:grow_dataflow}

\begin{figure}[t!] \centering
\includegraphics[width=0.485\textwidth]{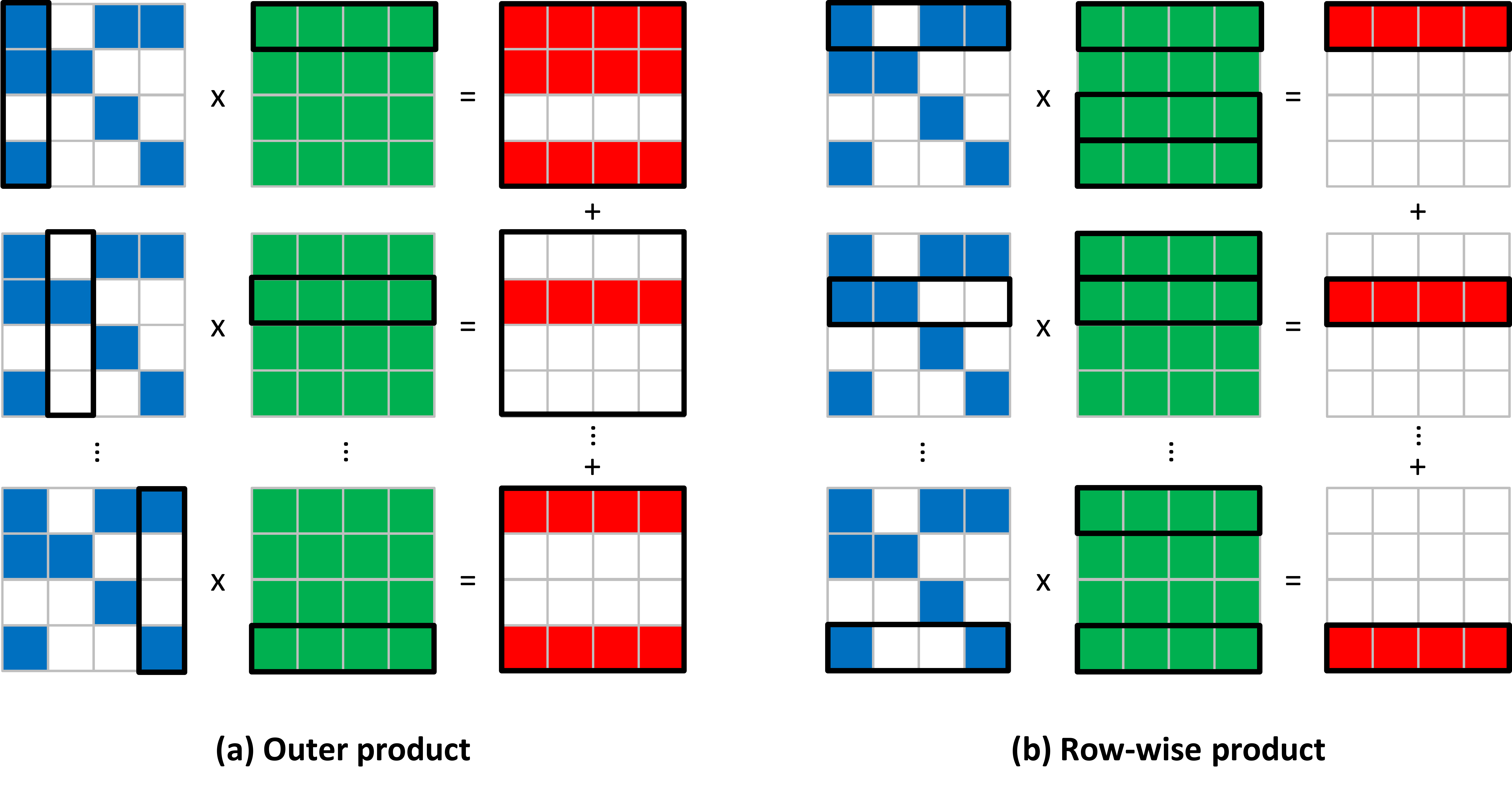}
\caption{
Outer-product vs. row-wise product GEMM dataflow..
}
\vspace{-.25em}
\label{fig:row_stationary}
\end{figure}

{\bf A row-stationary dataflow for \spgemm.}
Unlike GCNAX (or AWB-GCN) which employs an outer product (or inner product)
	based dataflow, \proposed employs a row-wise
	product GEMM dataflow based on Gustavson's algorithm~\cite{gustavson} (\fig{fig:row_stationary}).
In a row-wise product \spgemm dataflow, all the non-zero elements from
a single row of the left-hand side (LHS) matrix (the sparse, blue-colored elements)
are multiplied with the non-zero elements from the corresponding rows of the right-hand
side (RHS) matrix (the dense, green-colored elements), where the row index of
the RHS matrix is determined by the column index of the non-zero values of LHS matrix.
The results are accumulated into the corresponding row of the output matrix (the dense,
		red-colored elements). Because the $n$-th row in both the LHS matrix and the output 
matrix are all \emph{stationary} during the course of row-wise product, we henceforth
refer to this dataflow as \emph{row-stationary} to differentiate it from a pure
output-stationary dataflow~\cite{eyeriss}.

\begin{figure}[t!] \centering
\includegraphics[width=0.495\textwidth]{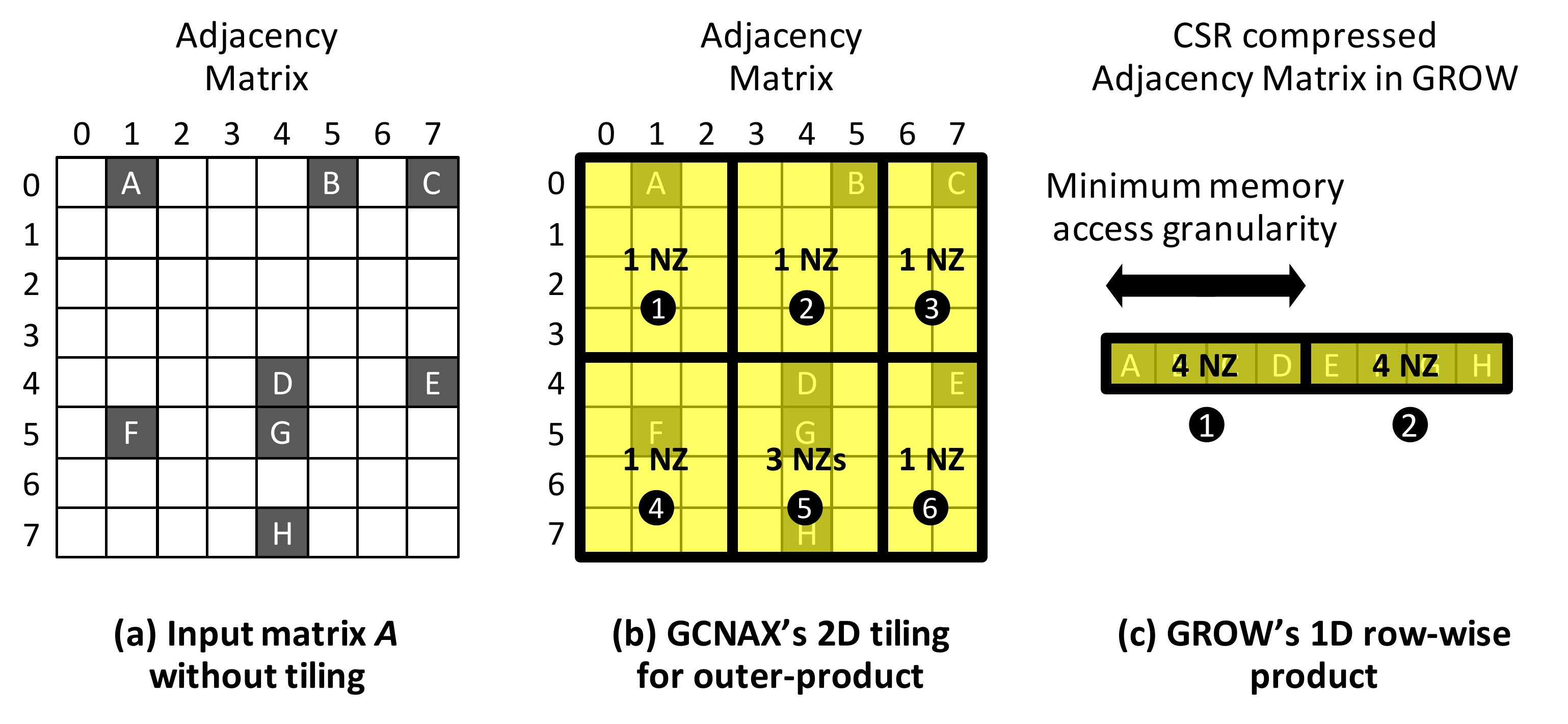}

\caption{
Difference between GCNAX's 2D tiling for outer-product and \proposed's 1D row-wise product.
The black numbered circles designate the order in which the input $2$D tile for (b) GCNAX
or the $1$D row for (c) \proposed is accessed for output derivation. (c) assumes the minimum
memory access granularity is four elements wide.
}
\vspace{-1.25em}
\label{fig:proposed_tiling}
\end{figure}

A key advantage of a row-wise product is twofold. First, multiple rows from the
	output matrix can be computed in parallel as there are \emph{no} data dependencies
	across different rows, allowing multiple processing engines to compute
	different rows of the output matrix without having to synchronize with each
	other. Because there are no dependencies across different rows, unlike the
	outer-product approach which mandates the entire ($2$D) input/output tiles to \emph{all} be
	stored on-chip, row-wise product enables a design 
	where several ($1$D) rows can be computed
	independently, allowing the required on-chip buffer space to store both the LHS
	matrix ($A$ and $X$) and the output matrix to be tuned to become relatively
	smaller than outer-product dataflow (yet with better utility). 
			 Such property  helps better utilize memory bandwidth when fetching the
			 active working set of the sparse matrix $A$.  \fig{fig:proposed_tiling}
			 compares the number of non-zero (NZ) elements available for processing
			 the \spgemm when employing GCNAX's 2D tiling strategy vs. \proposed's 1D
			 row-wise product. As discussed in \sect{sect:gcnax_bottleneck}
			 (\fig{fig:num_nz_in_tile}(a)), the number of non-zeros subject for
			 \spgemm within any given tile in GCNAX is typically less than $2$ in
			 aggregation. GCNAX, however, fails to accommodate such sparsity patterns
			 and applies a rigid 2D tiling window for both aggregation and
			 combination.  This leads to a significant waste in off-chip memory
			 bandwidth because the number of effectual bytes (i.e., the number of
					 non-zero elements within a given tile) fetched per $64$ bytes of
			 minimum memory access granularity becomes small
			 (\fig{fig:proposed_tiling}(b)).  Contrast that with \proposed's
			 row-stationary dataflow, where the number of effectual bytes fetched per
			 memory access and utilized for \spgemm is much higher and dense. This is
			 because the CSR compression format compacts all the non-zero elements
			 available within consecutive rows in a dense fashion
			 (\fig{fig:proposed_tiling}(c)), all of which will be
			 processed by our row-stationary \spgemm computation within a short time window.  Such
			 alignment of CSR compressed effectual data of matrix $A$ and the
			 row-wise product \spgemm leads to much higher off-chip bandwidth utility
			 as well as better on-chip buffer utilization.

{\bf Remaining challenges of row-stationary dataflow.} Nonetheless, a critical
trade-off made with the row-wise product (vs.  inner-/outer-product) is that
the RHS matrix ($XW$ and $W$) exhibits an \emph{irregular} reuse across the
derivation of \emph{different} output rows, the locality of which is strictly
dependent upon how the sparsity is manifested inside the LHS matrix (e.g., the
		$1$st row of the RHS matrix in \fig{fig:row_stationary}(b) is reused three
		times because there are three non-zero elements in the $1$st column of the
		LHS matrix). For the ($X$$\times$$W$) of combination stage, this is less of
a concern as the weight matrix $W$ is typically small in GCNs, allowing it to
be stored completely on-chip (\tab{tab:model_config}). Additionally, the
fraction of execution time the combination stage accounts for is relatively
small, especially for large-scale graph datasets
(\fig{fig:motivation_gcnax_latency_breakdown}). The aggregation stage
($A$$\times$$XW$), however, is a different story as the size of the RHS matrix
$A$ scales proportional to the number of graph nodes, the number of which can
amount to several millions and must be tiled across different iterations to be
temporally stored on-chip. Such property renders an intelligent caching
strategy that effectively balances both memory locality and parallelism
crucial. In the following subsections, we detail our software/hardware
co-design that maximizes locality (\sect{sect:clustering}) and parallelism
(\sect{sect:runahead}) for high efficiency.

%------------

\subsection{Enhancing Locality via Graph Partitioning}
\label{sect:clustering}

\begin{wrapfigure}{r}{0.6\linewidth}
\vspace{0.2em}
\includegraphics[width=0.985\linewidth]{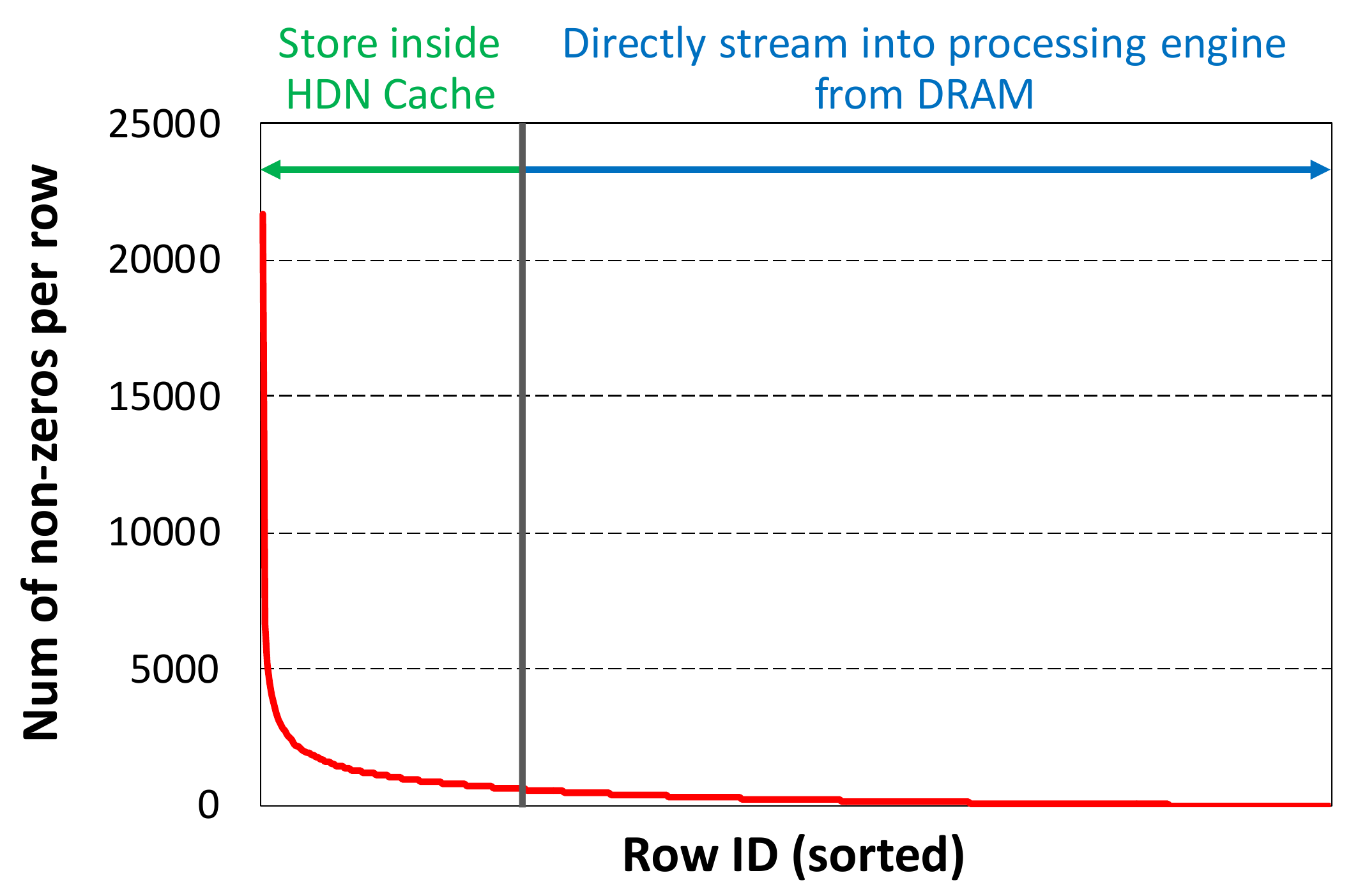}
\captionof{figure}{The power-law distribution of Reddit and how \proposed utilizes that property for caching HDNs.}
\vspace{0.5em}
\label{fig:power_law_and_locality}
\end{wrapfigure}

{\bf The power-law distribution in real-world graphs.}
It is well known that real-world graphs in critical application domains follow the \emph{power-law}
distribution (\fig{fig:power_law_and_locality}). Under the context of GCNs,
this implies that only a small number of rows (or columns) within
 the adjacency matrix $A$ accounts for the majority of non-zeros
 while the remaining majority of rows (or columns) only contain a
 few non-zeros.

\setlength{\columnsep}{8pt}%
\setlength{\intextsep}{0pt}%
\par

\begin{figure}[t!] \centering
\includegraphics[width=0.485\textwidth]{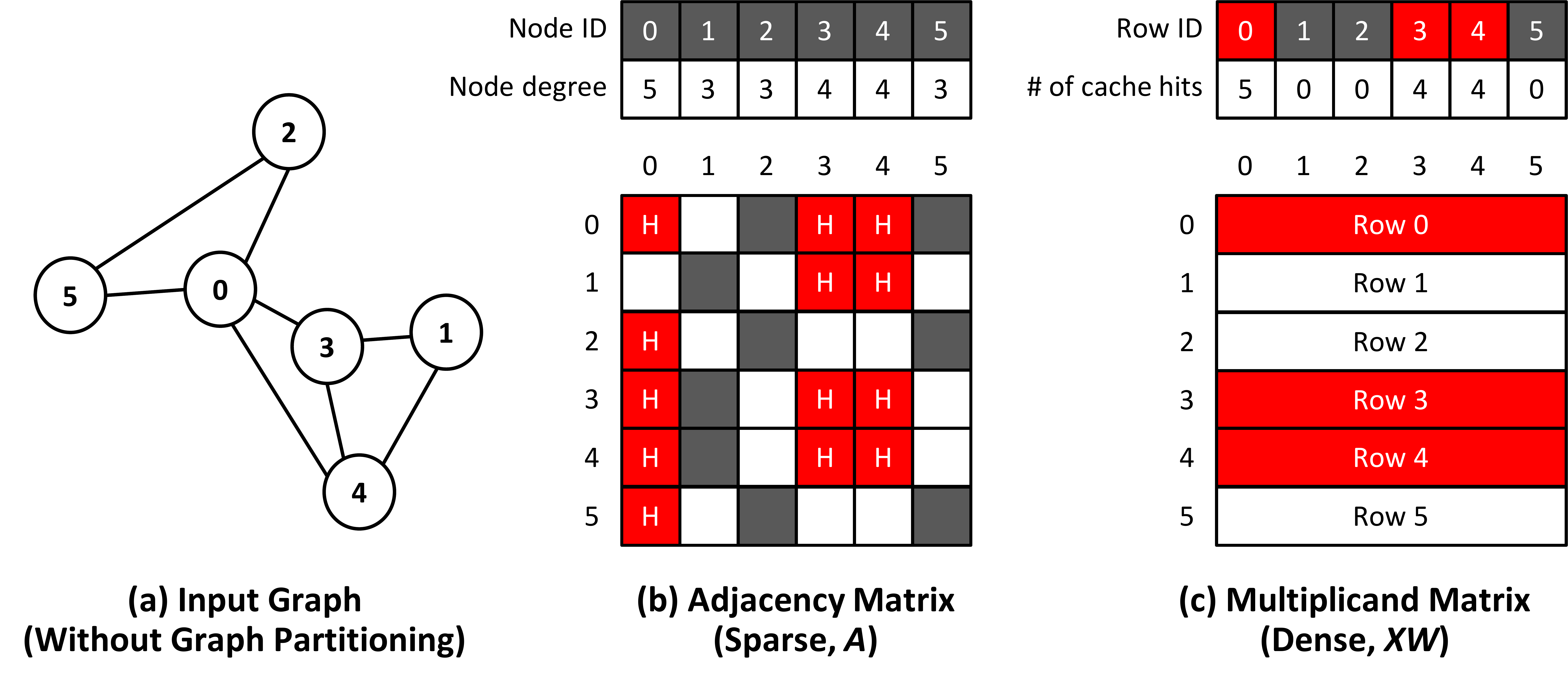}

\caption{
(a) Baseline input graph without graph partitioning and (b) its adjacency matrix. By only caching the top-$3$ high-degree nodes' RHS multiplicand matrix accesses (c), a total of $13$ HDN cache hits is achieved.
}
\vspace{-0.5em}
\label{fig:proposed_clustering_before}
\end{figure}

						 Our {\bf key approach} is to utilize the power-law distribution of
						 graph datasets to \emph{cost-effectively} capture the (high)
	locality of those  small number of HDN's memory accesses.  As discussed in
	\sect{sect:arch_overview}, the  \ibufdense contains 1) an \emph{HDN cache}
	that only captures the working set of high-degree node's RHS matrix accesses
	during aggregation and 2) an HDN ID list buffer that stores the list of
	top-$N$ HDN's IDs.  \fig{fig:proposed_clustering_before} illustrates an
	example of our approach, where nodes with ID$=$$0$,$3$,$4$ are determined as
	HDNs. The node IDs of the top-$N$ ($N=3$) high-degree nodes are stored as an
	array initially  inside DRAM, which the DMA unit fetches and forwards to the
	\ibufdense controller so that it stores the HDN's node ID information inside
	the HDN ID list buffer at the beginning of GCN inference (the details
			regarding \emph{how} the node IDs of HDNs are derived are discussed later
			in this subsection). The \ibufdense controller subsequently fetches the
	corresponding HDN's rows from the multiplicand matrix $XW$ and stores them
	inside the HDN cache.  Once the (CSR compressed) non-zeros of the target row
	of the adjacency matrix $A$ is fetched inside \ibufsparse, the processing
	engine starts executing the \spgemm, one row at a time (from the first to
			last row in \fig{fig:proposed_clustering_before}(b)).  During the
	derivation of the first output row, the \ibufdense controller examines the
	HDN ID list buffer against the column IDs of the non-zero values of the first
	row of the adjacency matrix, finding out that three out of the five $XW$ row
	requests can be serviced from the HDN cache (i.e., the red-colored ``H''
			(Hit) elements in the first row of \fig{fig:proposed_clustering_before}'s
			adjacency matrix). The controller therefore directly fetches the  three
	HDN's corresponding rows (ID$=$$0$,$3$,$4$) from the HDN cache while also
	requesting the DMA unit to stream in the (HDN cache missed) two low-degree node (LDN)'s rows
	(ID$=$$2$,$5$) from DRAM. 
	Overall, once the HDN cache is
	populated with rows $0$,$3$,$4$, \proposed can achieve a total of $13$ HDN
	cache hits while deriving the six output rows of $0-5$. 

\begin{figure}[t!] \centering
\includegraphics[width=0.485\textwidth]{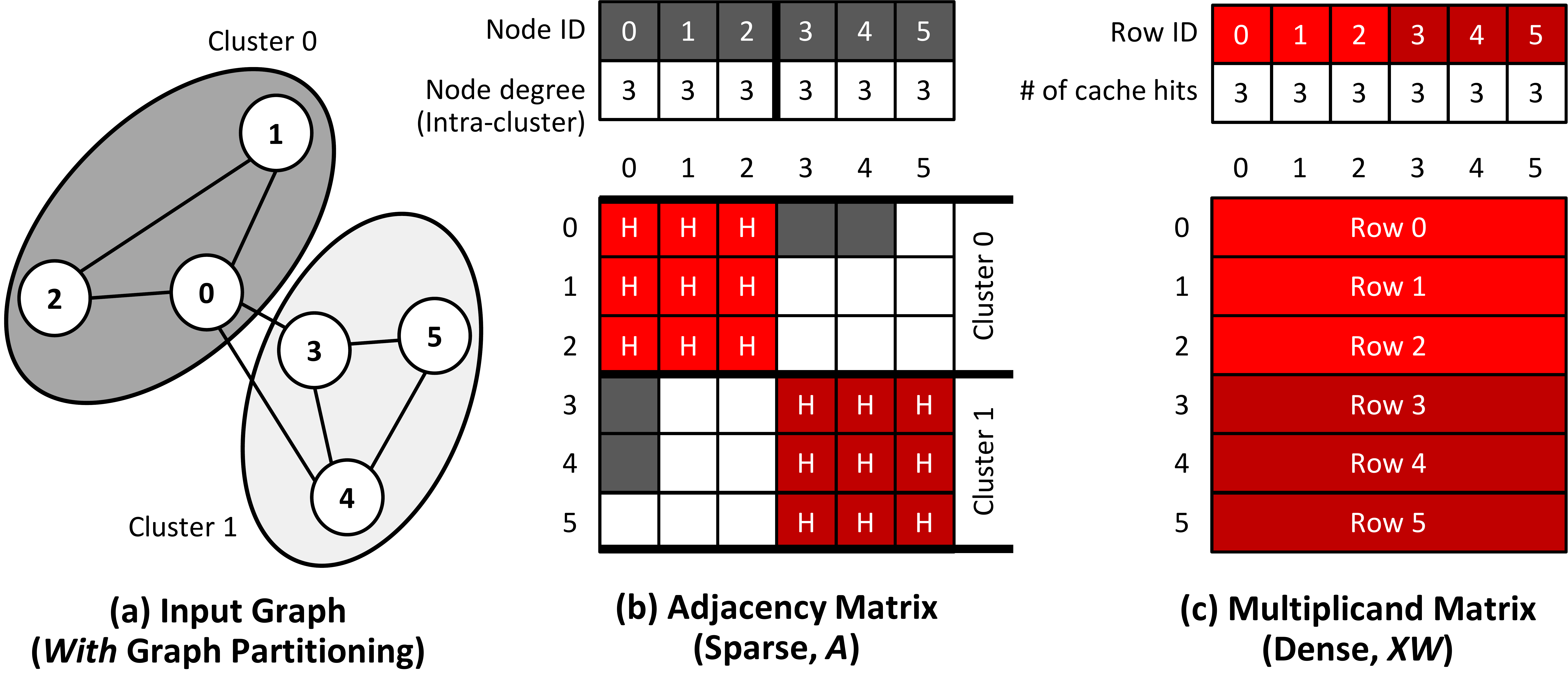}

\caption{
(a) Input graph \emph{with} graph partitioning (target of $2$ partitions, resulting in three nodes per cluster) and (b) its adjacency matrix. Graph partitioning helps increase the number of HDN cache hits to $18$.
}
\vspace{-1.25em}
\label{fig:proposed_clustering_after}
\end{figure}

{\bf Graph partitioning to maximize temporal locality.} As the HDN cache is
designed to capture the locality of only the HDN nodes \emph{without} accommodating
the usage of LDN nodes, one could consider the HDN cache as a \emph{scratchpad}
for HDN nodes. When the input graph is  small-scale, \proposed's scratchpad-like
cache microarchitecture effectively exploits the graph's power-law
distribution, achieving high HDN cache hit rate (up to $80\%$ for Cora,
		\sect{sect:eval_caching}) as well as significant speedup vs. GCNAX. Unfortunately,
	many real-world GCN applications like social network analysis or e-commerce
	are based on large-scale graph datasets, which contains an overwhelmingly
	large number of high-degree nodes for the statically sized HDN cache to
	sufficiently capture locality.  As such, the effectiveness of \proposed's
	caching mechanism lies in how to intelligently utilize the limited HDN cache
	space to sufficiently capture the overall locality inherent in the adjacency
	matrix $A$ as a whole, regardless of its size.

\begin{figure}[t!] \centering
\includegraphics[width=0.485\textwidth]{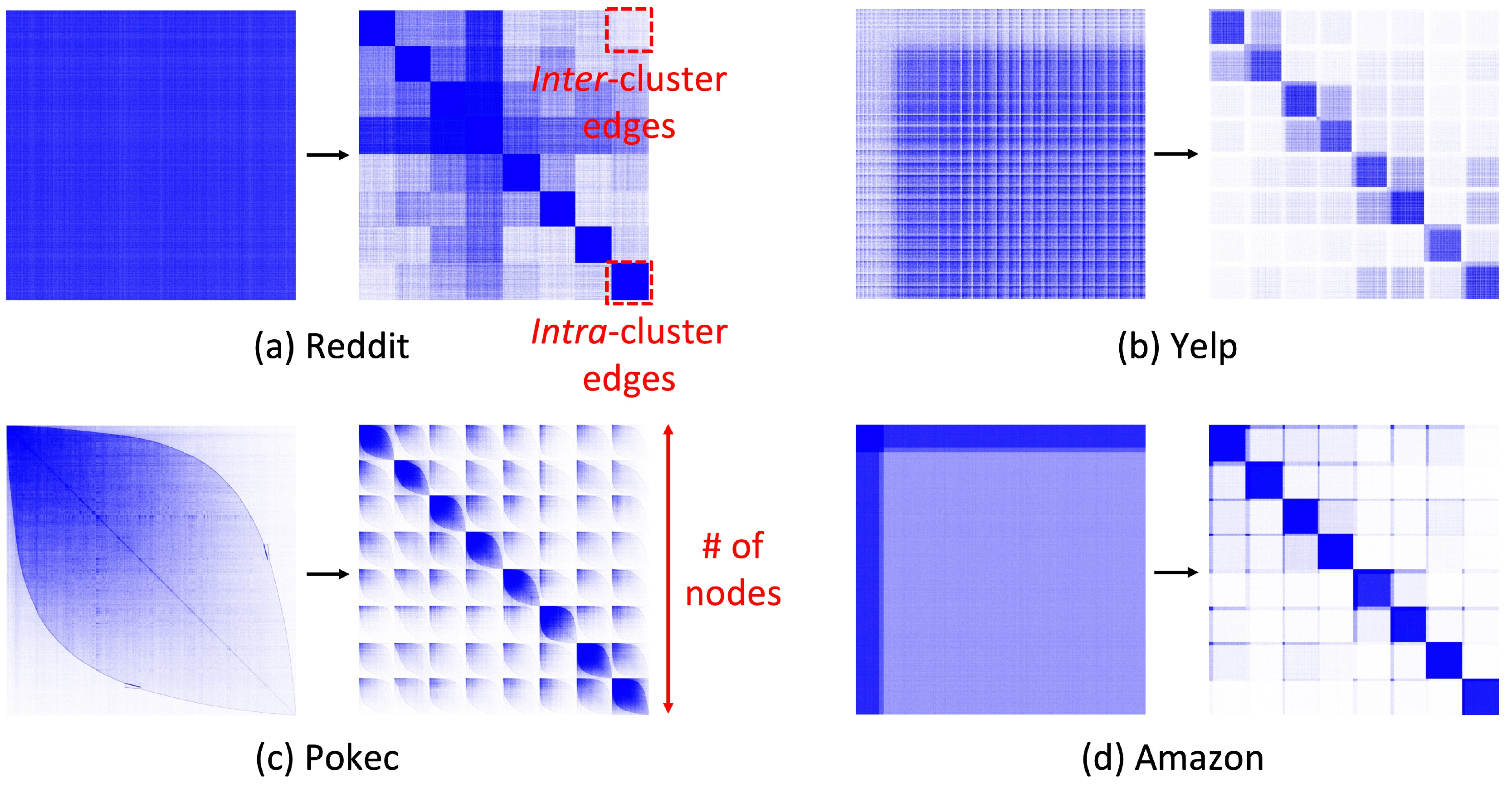}
\caption{
The effect of graph partitioning on the adjacency matrix of (a) Reddit, (b) Yelp, (c) Pokec, and (d) Amazon. For clear visualization, examples shown assume the graph partitioning algorithm targets $8$ output partitions, but our studied large-scale graphs are typically partitioned into thousands of clusters. Non-zero elements are colored in blue dots.
}
\vspace{-1.25em}
\label{fig:clustering_effect}
\end{figure}

To address such challenge, we use \emph{graph partitioning} algorithms to
partition the graph into multiple \emph{clusters}, allowing \proposed to
achieve high \emph{intra}-cluster temporal locality in GCN inference.  Graph
partitioning methods like Metis~\cite{metis} or Graclus~\cite{graclus} are designed to
construct the partitions over the input graph nodes such that
\emph{intra}-cluster nodes have much larger number of edges than
\emph{inter}-cluster nodes. In general, graph partitioning helps better capture
the clustering and community structure of the graph and is a widely employed
graph preprocessing technique for graph analytics.
\fig{fig:proposed_clustering_after} illustrates the effect of applying graph
partitioning algorithm on the input graph of
\fig{fig:proposed_clustering_before}(a). As depicted, graph partitioning only
changes the way a particular node is assigned with its node ID (i.e., node ID
		is changed from $1\rightarrow5$, $2\rightarrow1$, and $5\rightarrow2$ in
		\fig{fig:proposed_clustering_after}(a) vs.
		\fig{fig:proposed_clustering_before}(a)), 
				yet the effect it has on the
adjacency matrix is profound, especially for the purpose of \proposed's caching
strategy.  The key to our approach is to choose the top-N high-degree nodes
subject for HDN caching \emph{only within the cluster}, rather than across the
entire adjacency matrix. Consider the example shown in
\fig{fig:proposed_clustering_after}(b), where graph partitioning groups the
non-zero values of cluster $0$ (node ID$=0$,$1$,$2$) in the upper-left corner
of the adjacency matrix, while those of cluster $1$ are grouped at the
lower-right corner (node ID$=3$,$4$,$5$). By tracking the top-$N$ HDN nodes per
each cluster, \proposed can cache nodes ID=$0$,$1$,$2$ (and ID=$3$,$4$,$5$)
	during the first (and second) cluster's row-wise product based GEMM
	computation, leading to higher HDN cache hit rates than the baseline input
	graph. \fig{fig:clustering_effect} illustrates the effect of graph partitioning
	on a subset of our studied graph datasets, highlighting the several clusters of
	high-degree nodes around the diagonal line.
	In general, \proposed's HDN caching \emph{with} graph partitioning
	is shown to achieve significant improvements in HDN cache hit rates (up to $17\times$ higher), which we quantify later in \sect{sect:eval_caching}.

\begin{figure*}[t!] \centering
\includegraphics[width=0.98\textwidth]{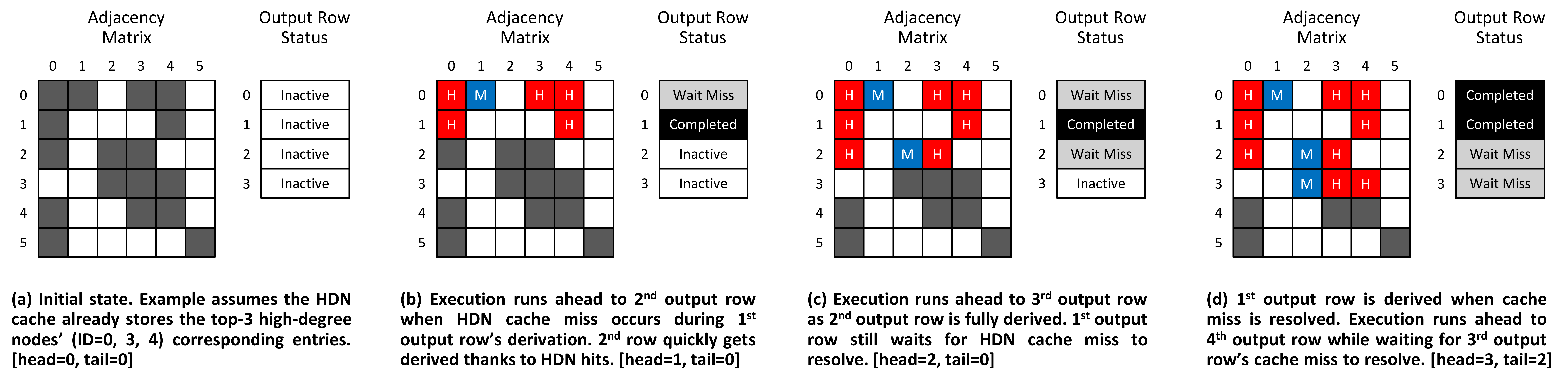}

\caption{
An example illustrating \proposed's multi-row stationary runahead execution mode.
}
\vspace{-1.25em}
\label{fig:proposed_runahead}
\end{figure*}

{\bf Design overhead.} Unlike the GCN input matrix $X$ whose value can change
over different inputs, both the values as well as the structure of the
adjacency matrix $A$ is statically fixed, regardless of what the GCN input is.
Our proposal is to \emph{preprocess} the adjacency matrix $A$ using graph
partitioning, thereby generating multiple clusters of graph partitions with
high temporal locality.  Our software stack augments the Metis graph
partitioning algorithm~\cite{metis} with a pass that generates the top-$N$ high-degree nodes
as a HDN ID list per each cluster, allowing the runtime hardware controller to
fetch the HDN ID list before a target cluster begins computation. 
For a given target input graph, \proposed's graph
preprocessing step incurs a one-time latency cost, as the \emph{partitioned}
graph and its HDN ID lists are used as-is for all future inference runs without modification.
Consequently. the latency overheads of GROW graph preprocessing step
(ranging from tens of milliseconds to several tens of minutes depending
		on the number of graph nodes) is amortized over all future GCN inference runs.
In terms of storage overhead, the default \proposed configuration employs a
$12$ KB SRAM buffer within the \ibufdense in order to store the $4096$ HDN ID list
of the current cluster (i.e., $4096$$\times$$3$B, $3$ Bytes per ID). The entire
HDN ID lists for all the clusters per each graph amounts to several MBs of storage
($12$ KB per cluster, up to
		several thousands of clusters per graph) which is kept in DRAM.

\subsection{Enhancing Parallelism via Runahead Execution}
\label{sect:runahead}

{\bf Motivation.} While \proposed's graph partitioning helps significantly
reduce the number of HDN cache misses, the likelihood of any given output row's
derivation to incur a HDN cache miss is still high. 
We observe that, except for the small-scale Cora and Citeseer, an average
$81\%$ of the output rows experience more than a single HDN cache miss
per each row's derivation
(i.e., some of its HDN cache queries lead to hits while others result
		in misses).
Having \proposed \emph{actively} work on just a single output row
would therefore be highly suboptimal as the latency to service the HDN cache
misses would directly be exposed to the processing engine, effectively blocking
the derivation of \emph{other} output rows and causing severe performance
overheads.

{\bf Runahead execution with multi-row stationary dataflow.} To address the
aforementioned research challenge, \proposed employs a \emph{multi-row
	stationary runahead execution} scheme as means to maximize memory-level
	parallelism and hide the latency of HDN cache misses.
The design objective of \proposed's multi-row stationary dataflow is to concurrently work
on the derivation of multiple output rows, which is achieved by 
provisioning both the \ibufsparse and  \obuf
	buffers to be large enough to keep track of multiple output rows' derivation process.
	\fig{fig:proposed_runahead} is an example of \proposed multi-row stationary
	runahead execution assuming a $4$-way multi-row window (i.e., up to $4$
			output rows can be actively worked upon by the processing engine) with
	node ID$=$$0$,$3$,$4$ determined as the HDN ID list.  While executing the
	first output row, the \proposed control unit experiences a HDN cache miss.
	Rather than idly waiting for the HDN cache miss to be resolved, the control
	unit runs ahead to the second row by fetching the second row's (compressed)
	non-zero values of the adjacency matrix (\fig{fig:proposed_runahead}(b)).
	Because the two non-zero elements of $A$'s second row are all HDN cache hits,
	derivation of the second output row can be completed shortly while waiting
	for the first row's cache miss to be serviced. Assuming the first row's cache
	miss is not serviced in a timely manner, even though the second row has already completed
	its execution, \proposed can keep running ahead 
	to the third output
	row's derivation (\fig{fig:proposed_runahead}(c)), further hiding the
	latency penalty of the first row's HDN cache miss. While waiting for the third
	row's cache miss to be resolved, \proposed again runs ahead to the fourth
	row's processing (\fig{fig:proposed_runahead}(d)), kicking off two more HDN cache misses inflight, maximizing
	memory-level parallelism and hiding its latency.

\begin{figure}[t!] \centering
\includegraphics[width=0.485\textwidth]{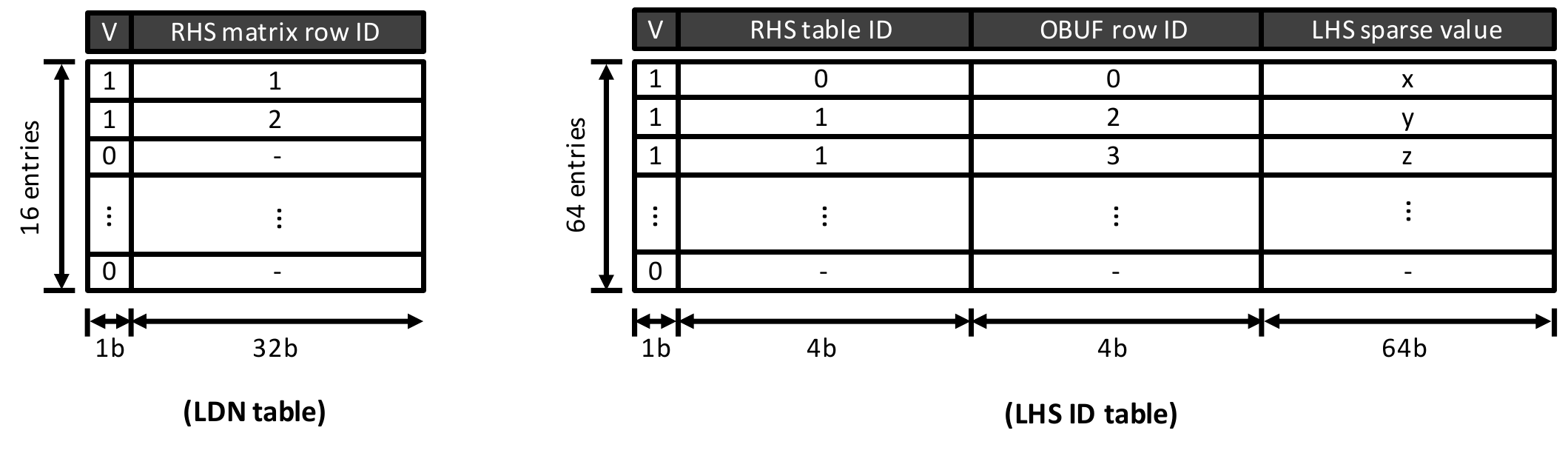}
\caption{
GROW microarchitectures enabling multi-row stationary runahead execution. The status of the two tables
	is updated assuming the same example shown in \fig{fig:proposed_runahead}, where LDN nodes with ID=$1$,$2$
	miss in the HDN cache, allocating two entries inside the LDN table. Because output rows $0$, $2$, and $3$
	requires RHS matrix rows of ID=$1$,$2$, this information is tracked inside the LHS ID table using three
	separate entries (the LHS sparse values to be multiplied with these three rows are assumed as $x$, $y$, $z$).
}
\vspace{-1.2em}
\label{fig:revision_mshr}
\end{figure}

{\bf Implementation overhead.}
GROW's runahead execution requires an MSHR (miss-status holding register) like microarchitecture that
keeps track of which HDN cache queries have missed so far (i.e., LDN nodes) and are under the process of being fetched from the memory
subsystem.
 \fig{fig:revision_mshr}
provides a high-level overview of the key microarchitectural components that enable GROW's runahead execution. First, an $M$
entry \emph{LDN table} keeps track of the LDN nodes that missed in the HDN cache and is in need to fetch the missed RHS matrix rows
from DRAM. Second, another
$N$ entry \emph{LHS ID table} stores the non-zero (sparse) LHS matrix values to be multiplied with the
(HDN cache missed) RHS matrix rows. Once the missed RHS matrix row is fetched on-chip, the returned row's
corresponding table index ID allocated in the LDN table is used to conduct a CAM lookup inside the LHS ID table,
allowing the corresponding O-BUF$_{dense}$ to be updated properly. We empirically find that having these two tables
be sized at $M=16$, $N=64$ sufficiently captures the benefits of runahead execution, amounting to $64$ Bytes (LDN table)
and $544$ Bytes (LHS ID table) of storage space, respectively. Such design overhead is amortized over GROW's several hundreds of KBs
worth of on-chip SRAM capacity (detailed in 
\tab{tab:grow_area}).

%% file: tex/methodology.tex
\section{Evaluation Methodology}
\label{sect:methodology}

\sethlcolor{yellow}
{\bf Performance.} We model \emph{both} GCNAX and \proposed as a
cycle-level simulator implemented using C++. The simulator is driven by the studied graph datasets
and the GCN input feature matrices (\tab{tab:model_config}) which we extracted
using PyTorch Geometric~\cite{pyg}, SNAP (Stanford Network Analysis
		Project)~\cite{snapdataset}, and OGB (Open Graph Benchmark)~\cite{ogb}.
For a fair comparison against state-of-the-art GCNAX, \proposed has been
configured to provide  the same level of computation throughput and off-chip
memory bandwidth  while provisioned with similar on-chip SRAM capacity.
\tab{tab:grow_parameters} summarizes the key architectural parameters of
\proposed's baseline configuration.  

{\bf Area.} We measure \proposed's area by implementing it in RTL using SystemVerilog.
The RTL model is synthesized with Synopsys
Design Compiler targeting $1$ GHz of operating frequency using a $65$ nm
standard-cell library. 
The largest sized HDN cache is designed using $16$ banks of
SRAM arrays, each array synthesized as a single-ported SRAM using a memory compiler. The fully associative
HDN ID list is designed and synthesized as a CAM using D-flipflops for maximal performance, allowing
a single cache lookup per each clock cycle. The small sized \ibufsparse and \obuf are implemented
using dual-ported SRAMs and D-flipflops, respectively.
GCNAX reports its area numbers under a $40$ nm technology,
	so we scale our area estimations from our $65$ nm results and report 
	estimated numbers for $40$ nm	when comparing against GCNAX (\tab{tab:grow_area}).

{\bf Energy.} When quantifying energy consumption, we employ the energy model from \cite{horowitz:isscc}
for quantifying energy per operation for both arithmetic operations as well as off-chip DRAM accesses.
For modeling the power and energy consumption of on-chip SRAM usage, we use CACTI~\cite{cacti} targeting
a $45$nm process. Since the area of both GCNAX and \proposed is dominated by 
on-chip SRAM buffer space,
	we use CACTI's leakage power to estimate static energy consumption.

\begin{table}[tpb]
    \centering
    
    \caption{\proposed architecture configuration.}
    
	\scriptsize
    \begin{tabular}{|l|c|}
    \hline
    \multicolumn{1}{|c|}{{\textbf{Parameter}}} & \textbf{Value} \\
    \hline
    MAC width & $64$ bits \\
    \# MACs & $16$ \\
    \ibufsparse & $12$ KB \\
    HDN ID list & $12$ KB \\
    HDN cache & $512$ KB \\
    \obuf  & $2$ KB \\
    Runahead execution degree & $16$\\
    Memory bandwidth & $128$ GB/sec\\

    \hline
    \end{tabular}
    \vspace{-.5em}
    \label{tab:grow_parameters}
\end{table}

\begin{table}[tpb]
\centering
\caption{GROW vs. GCNAX area breakdown.}

\scriptsize
\begin{tabular}{|l|c|c|c|}
\hline
\multirow{3}{*}{\bf{Component}} & \multicolumn{3}{c|}{\bf{Area ($mm^2$)}}\\
\cline{2-4}
 & $40$ nm & $40$ nm & $65$ nm\\
 & (GCNAX) & $(estimated)$ & $(measured)$\\
\hline
\hline
MAC array & - & 0.232 & 0.613\\
\ibufsparse & - & 0.121 & 0.319\\
HDN ID list & - & 0.421 & 1.112\\
HDN cache & - & 1.352 & 3.569\\
\obuf & - & 0.043 & 0.113\\
Others & - & 0.022 & 0.059\\
\hline
\hline
Total & 6.51 & 2.191 & 5.785 \\
\hline
\end{tabular}
\vspace{-.5em}
\label{tab:grow_area}
\end{table}

%% file: tex/evaluation.tex
\section {Evaluation} 
\label{sect:evaluation}

\subsection{Caching Efficiency and Memory Bandwidth Usage}
\label{sect:eval_caching}

\begin{figure}[t!] \centering
\includegraphics[width=0.485\textwidth]{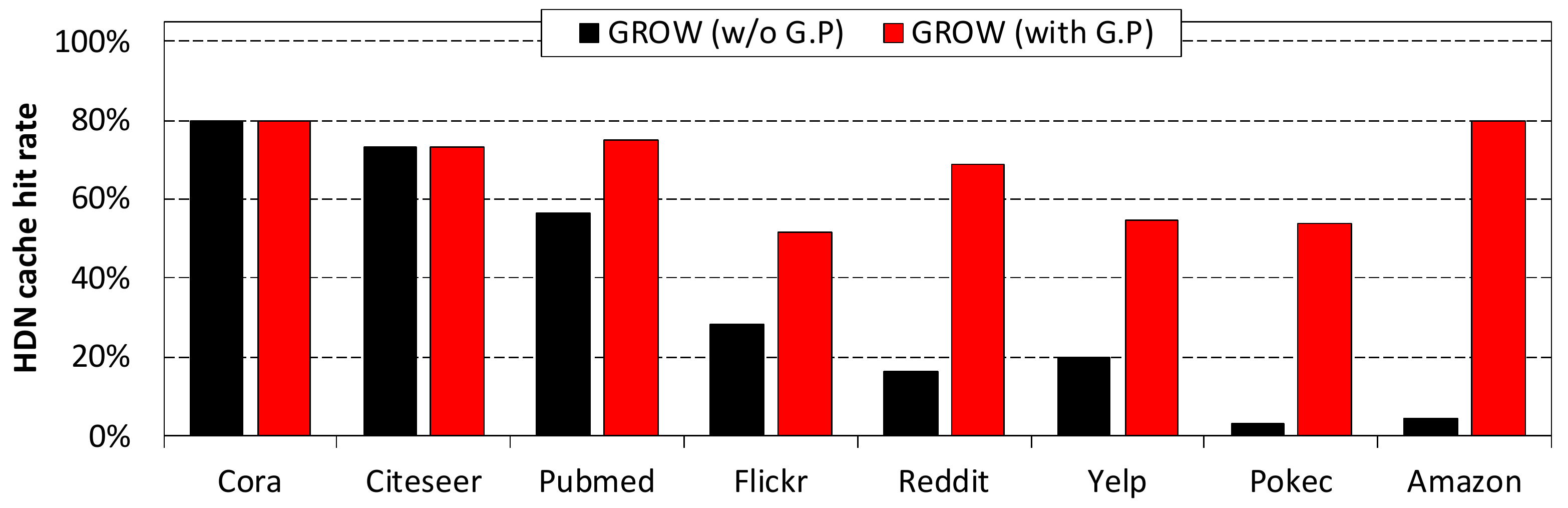}

\caption{
HDN cache hit rate with and without graph partitioning (G.P). Caching \emph{without} graph partitioning simply caches the top-$N$ (=$4096$) high-degree nodes.
}
\vspace{-.2em}
\label{fig:eval_eval_hdn_hit_rate}
\end{figure}

\begin{figure}[t!] \centering
\includegraphics[width=0.485\textwidth]{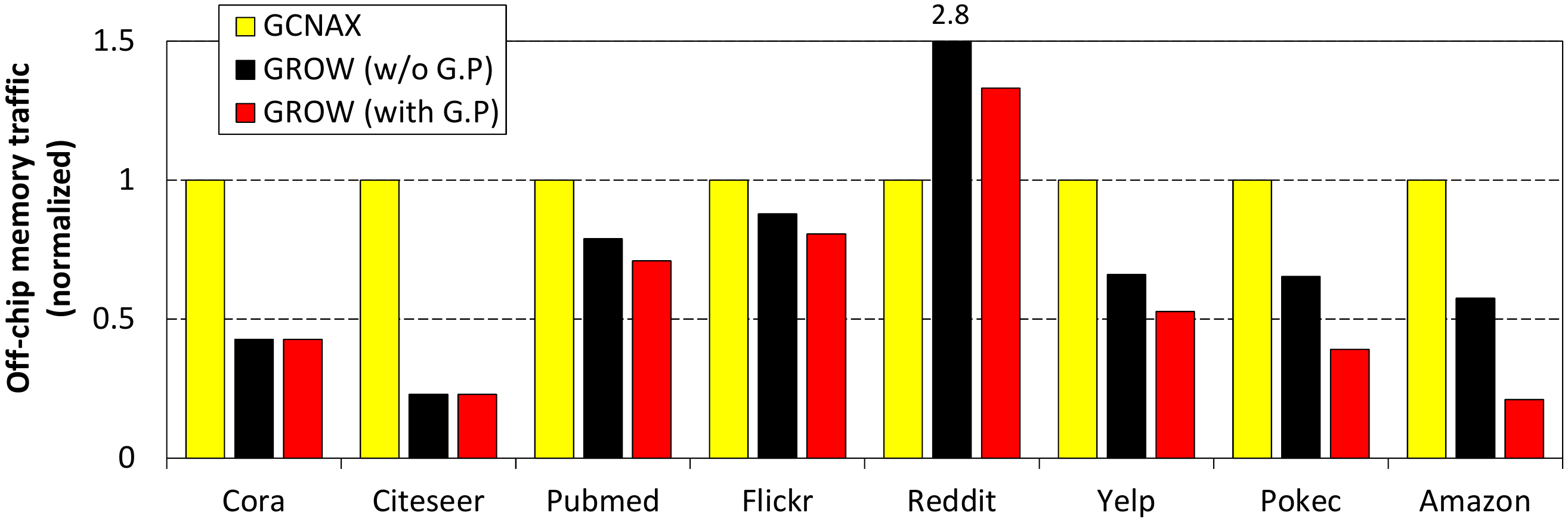}
\caption{
Total number of bytes read from DRAM (normalized to GCNAX).
}
\vspace{-.2em}
\label{fig:eval_dram_traffic}
\end{figure}

	 \fig{fig:eval_eval_hdn_hit_rate} compares \proposed's HDN cache hit rate
	 with and without graph partitioning (denoted G.P). For small-scale graphs
	 like Cora and Citeseer, the HDN cache is able to stash the majority of dense
	 rows with high locality (all of the HDN cache misses are compulsory misses that the DMA
			 unit brings on-chip immediately), leading to high HDN cache hit rate regardless
	 of graph partitioning being employed.
	 For large-scale graph datasets, however, graph partitioning
	 provides substantial improvements in HDN hit rates (e.g.,
			 $5\%$$\rightarrow$$79\%$ for Amazon) which helps \proposed dramatically
	 reduce the off-chip memory traffic as illustrated in
	 \fig{fig:eval_dram_traffic}.  Reddit exhibits an unusually high average node
	 degree (\tab{tab:model_config}, \fig{fig:clustering_effect}(a)), which
	 enables GCNAX to better exploit locality when conducting the
	 $2$ planar tiled outer-product \spgemm, having \proposed incur $31\%$
	 higher memory traffic.  Nonetheless, \proposed with graph partitioning shows
	 high robustness by significantly reducing off-chip memory accesses
	 consistently across the other graph datasets, achieving $2\times$
	 (max $4.7\times$) reduction in DRAM traffic vs. GCNAX.

\begin{figure}[t!] \centering
\includegraphics[width=0.485\textwidth]{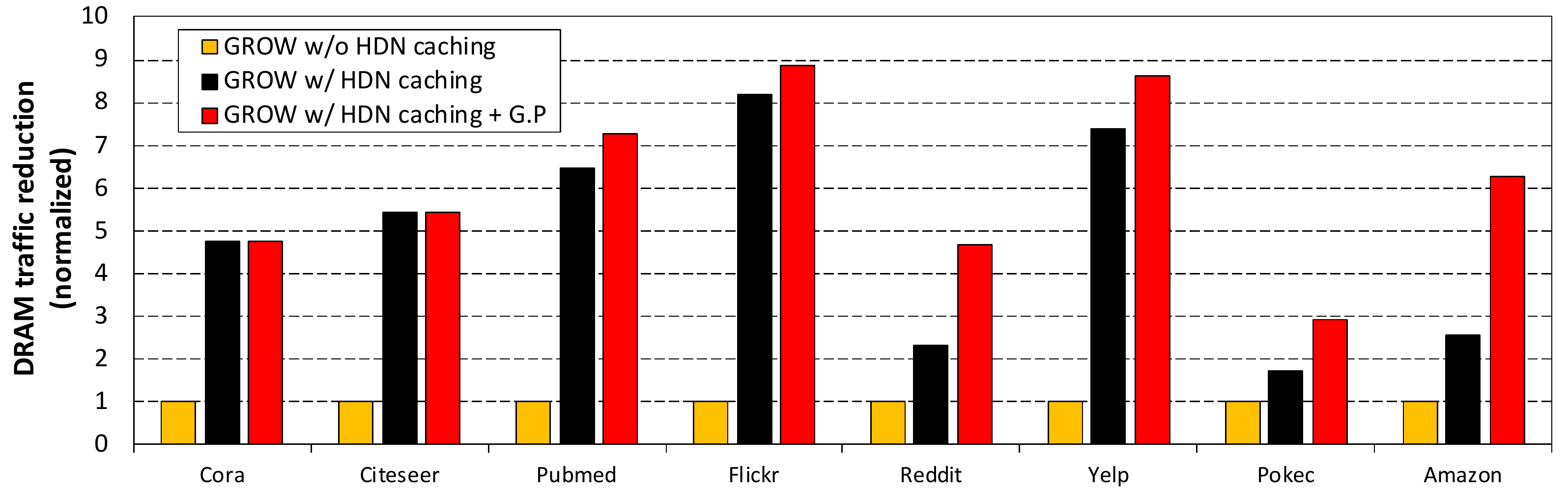}
\caption{
The effect of HDN caching and graph partitioning on GROW's DRAM traffic reduction (normalized 
		to GROW without HDN caching and graph partitioning). The higher the better.
}
\vspace{-1em}
\label{fig:revision_caching_effect}
\end{figure}

To better isolate the effect of HDN caching with graph partitioning, we also show in
\fig{fig:revision_caching_effect}
the memory traffic reduction achieved with and without HDN caching and graph partitioning. 
GROW without HDN caching and graph partitioning incurs a significantly higher DRAM burden, incurring an
average $4.3\times$ higher memory traffic than GROW with HDN caching but without graph partitioning ($5.8\times$
		higher traffic than GROW with both HDN caching and graph partitioning).

\subsection{Performance}
\label{sect:eval_perf}

\begin{figure}[t!] 
\centering
\subfloat[]{
\includegraphics[width=0.485\textwidth]{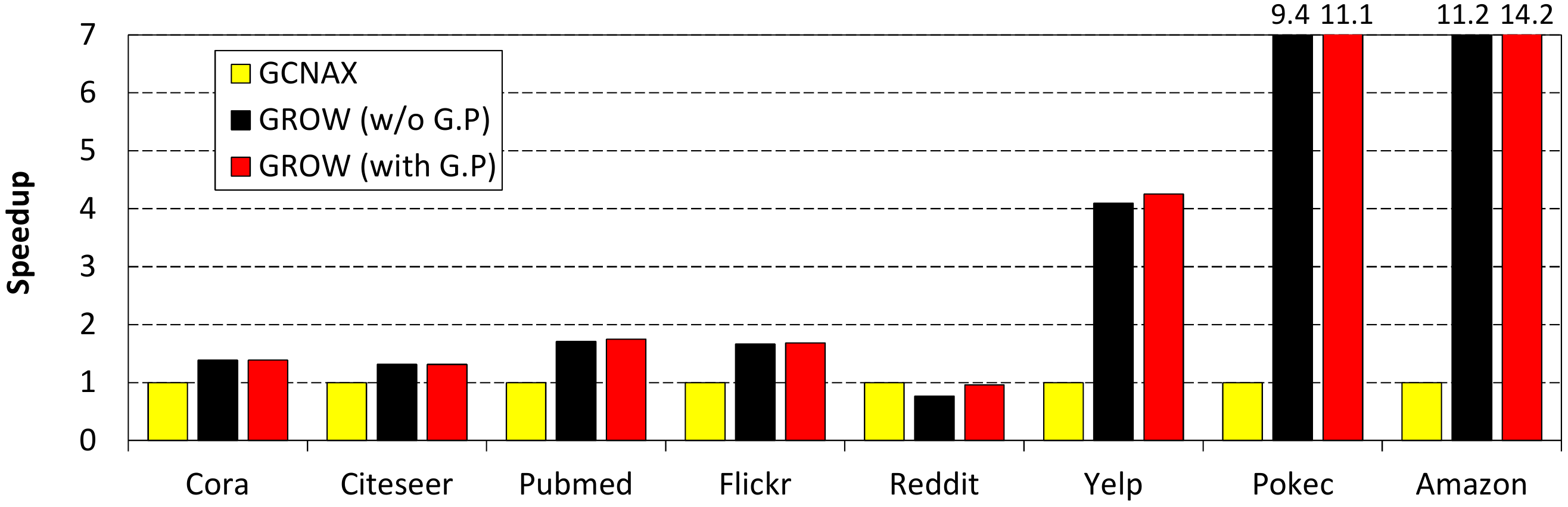}
%\vspace{-.2em}
	\label{fig:}
}
\vspace{0.5em}
\subfloat[]{
	\includegraphics[width=0.485\textwidth]{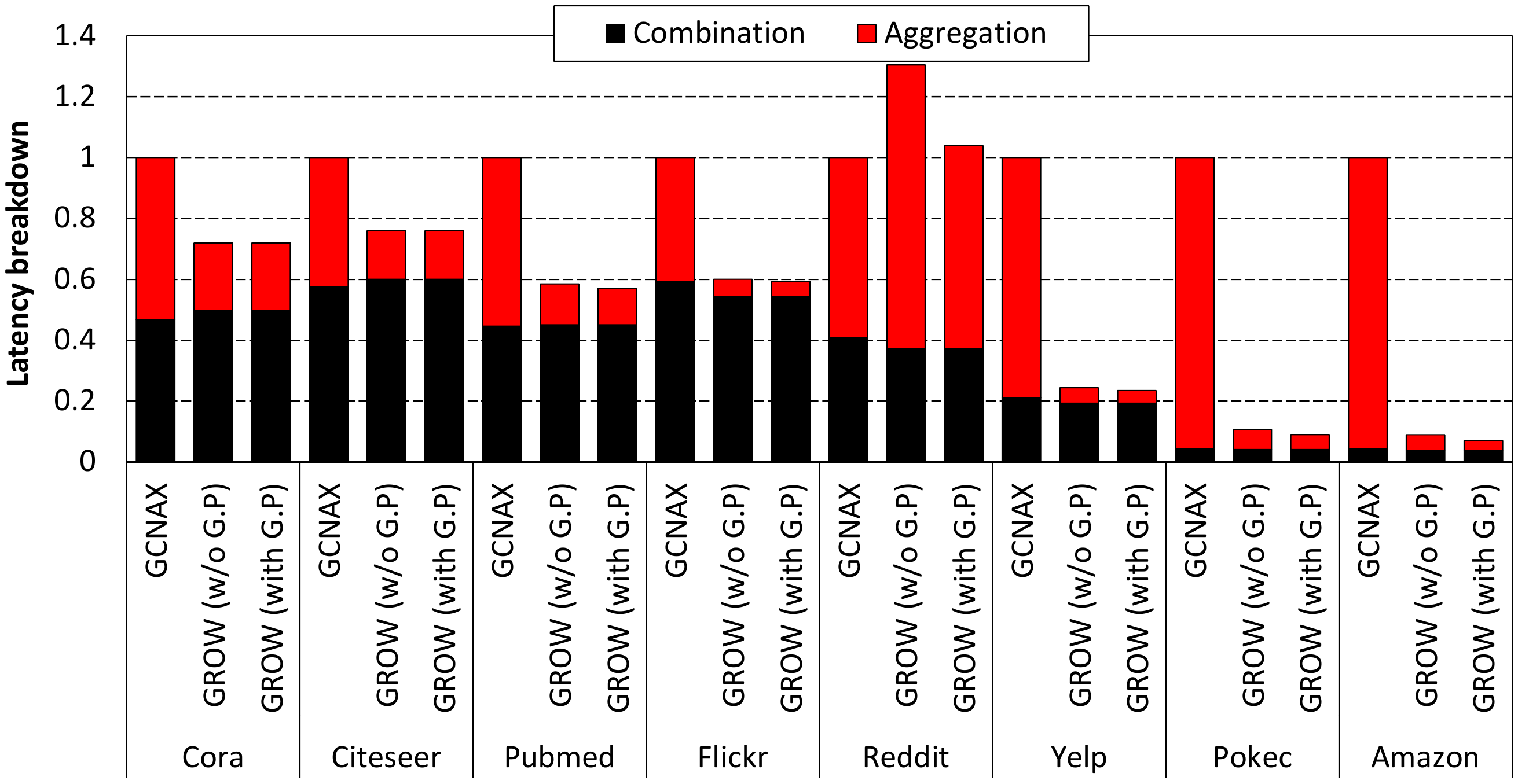}
%	\vspace{-.2em}
	\label{fig:}
}

\caption{ 
(a) \proposed speedup vs. GCNAX and (b) a latency breakdown (normalized to GCNAX).}   
\vspace{-.2em}
\label{fig:eval_speedup}
\end{figure}

\proposed significantly improves the performance of GCNs, especially for the
large-scale and sparse graph datasets, achieving an average $2.8\times$
speedup (max $14.2\times$) vs. GCNAX. 
It is worth pointing out that the bulk of \proposed's performance benefits comes from
greatly reducing the latency spent in conducting the aggregation stage. As discussed  in
\fig{fig:motivation_gcnax_latency_breakdown}, GCNAX spends significant fraction of execution
time in the aggregation stage, especially for the large-scale graphs like Yelp, Pokec, and Amazon.
 As depicted in \fig{fig:eval_speedup}(b), 
\proposed successfully reduces the latency in this critical bottleneck stage
by an average $6.3\times$, 
 shifting the GCN inference bottleneck now to the combination stage.

\begin{figure}[t!] \centering
\includegraphics[width=0.47\textwidth]{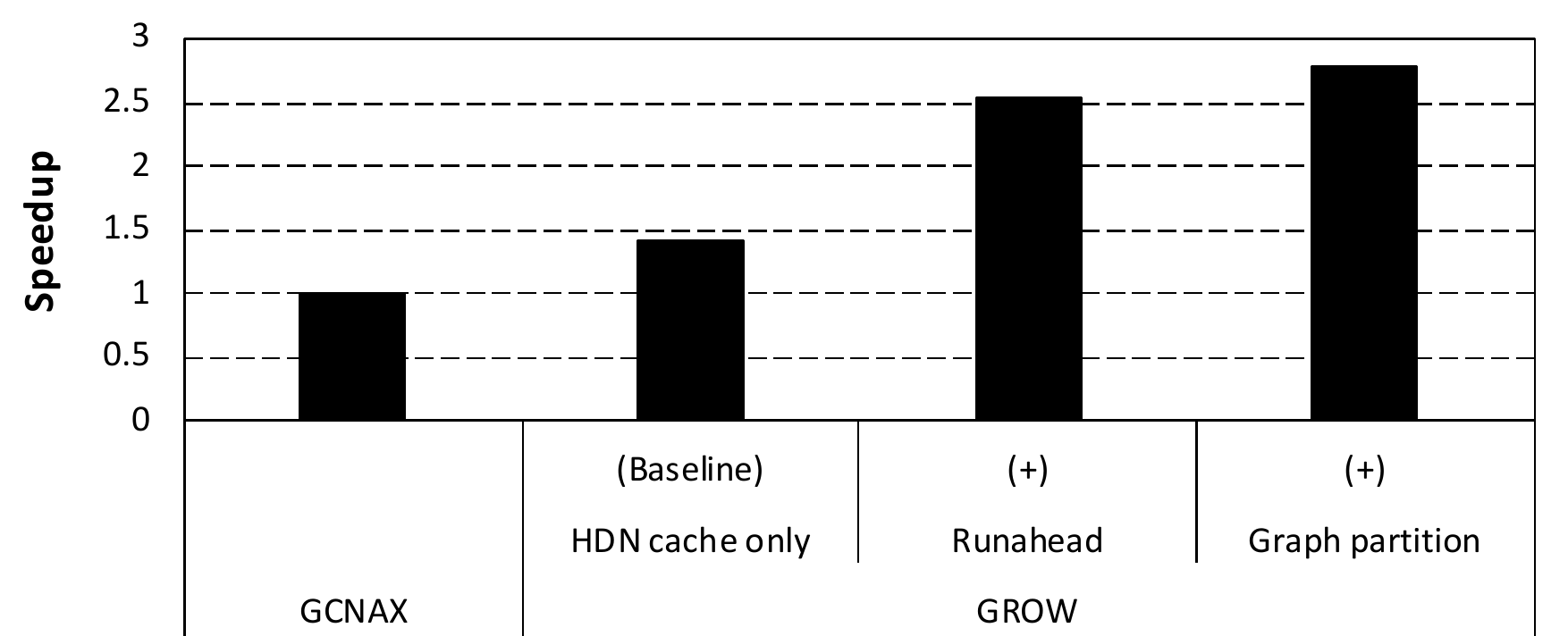}
\caption{
GROW's average speedup when incrementally applying our proposed optimizations one-by-one (from left to right).
}
\vspace{-1em}
\label{fig:eval_ablation}
\end{figure}

\subsection{Ablation Study}
\label{sect:eval_ablation}

To clearly demonstrate where GROW's speedup comes from, we quantify the impact
	of isolating our three main proposals in \fig{fig:eval_ablation} 
	as an 
	ablation study. The baseline GROW employs the proposed row-stationary dataflow
		with HDN caching but without runahead execution nor graph partitioning, achieving
an average $1.4\times$ speedup. Applying runahead execution provides a further $1.8\times$
speedup, with graph partitioning additionally reducing latency by $1.1\times$.

\subsection{Energy-Efficiency}
\label{sect:eval_energy}

\begin{figure}[t!] \centering
\includegraphics[width=0.485\textwidth]{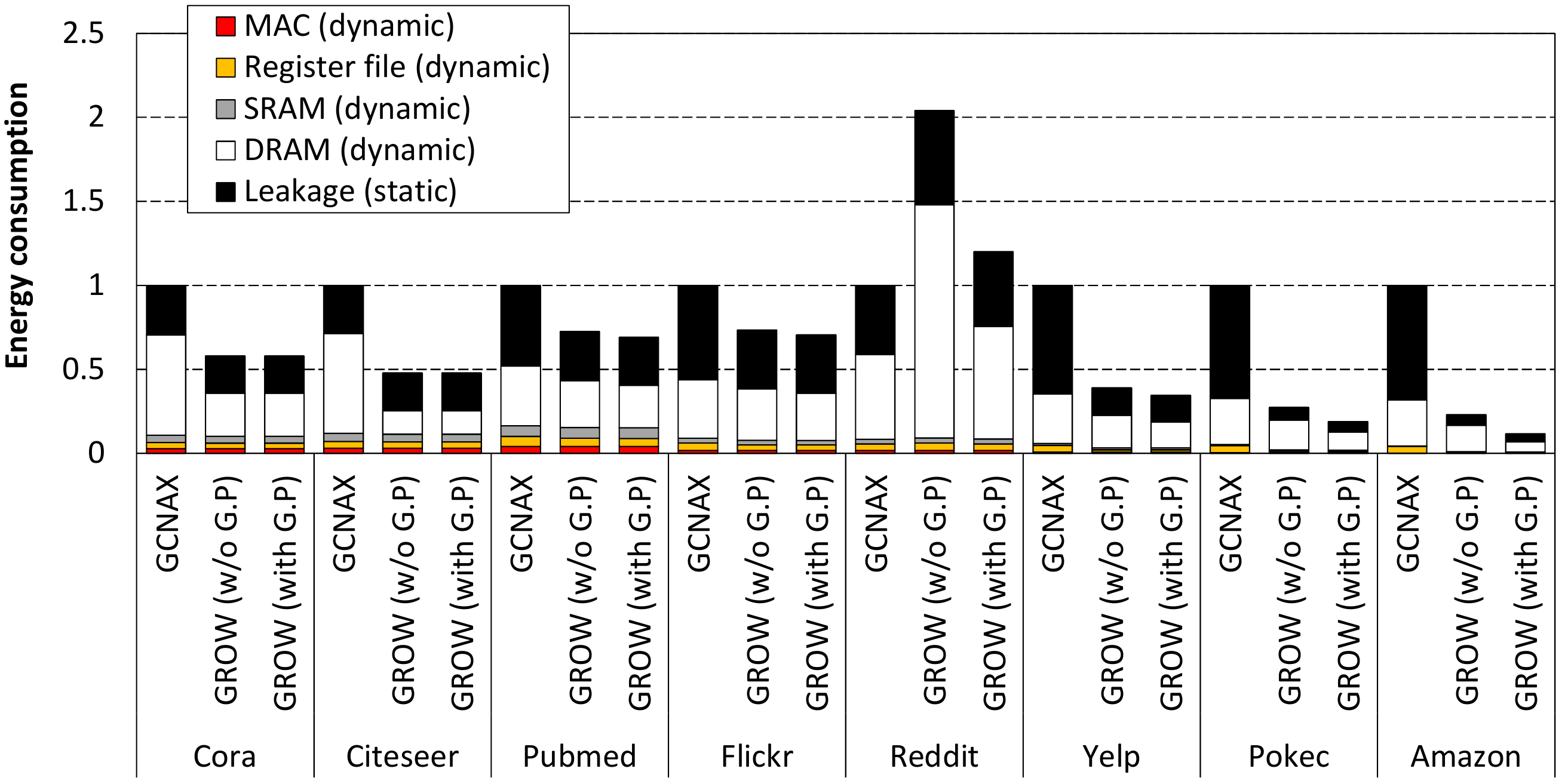}

\caption{
Energy consumption of \proposed (normalized to GCNAX).
}
\vspace{-1.25em}
\label{fig:eval_energy_efficiency}
\end{figure}

\fig{fig:eval_energy_efficiency} shows a breakdown of energy consumption in
GCNAX and the two design points of our \proposed architecture. Given the memory
bound nature of \spgemm, a large portion of dynamic energy is consumed in off-chip data
movements rather than on-chip SRAM accesses or compute.
Thanks to the significant reduction in memory traffic, \proposed
noticeably reduces energy consumed in off-chip memory accesses. Additionally,
					 \proposed also significantly reduces static energy consumption
					 thanks to the reduction in execution time, achieving an overall
					 $2.3\times$ improvement in energy-efficiency.
 
\subsection{Area Analysis}
\label{sect:eval_area}

\setlength{\columnsep}{8pt}
\setlength{\intextsep}{0pt}
\begin{wrapfigure}{r}{0.485\linewidth}
\vspace{0.2em}
\includegraphics[width=0.98\linewidth]{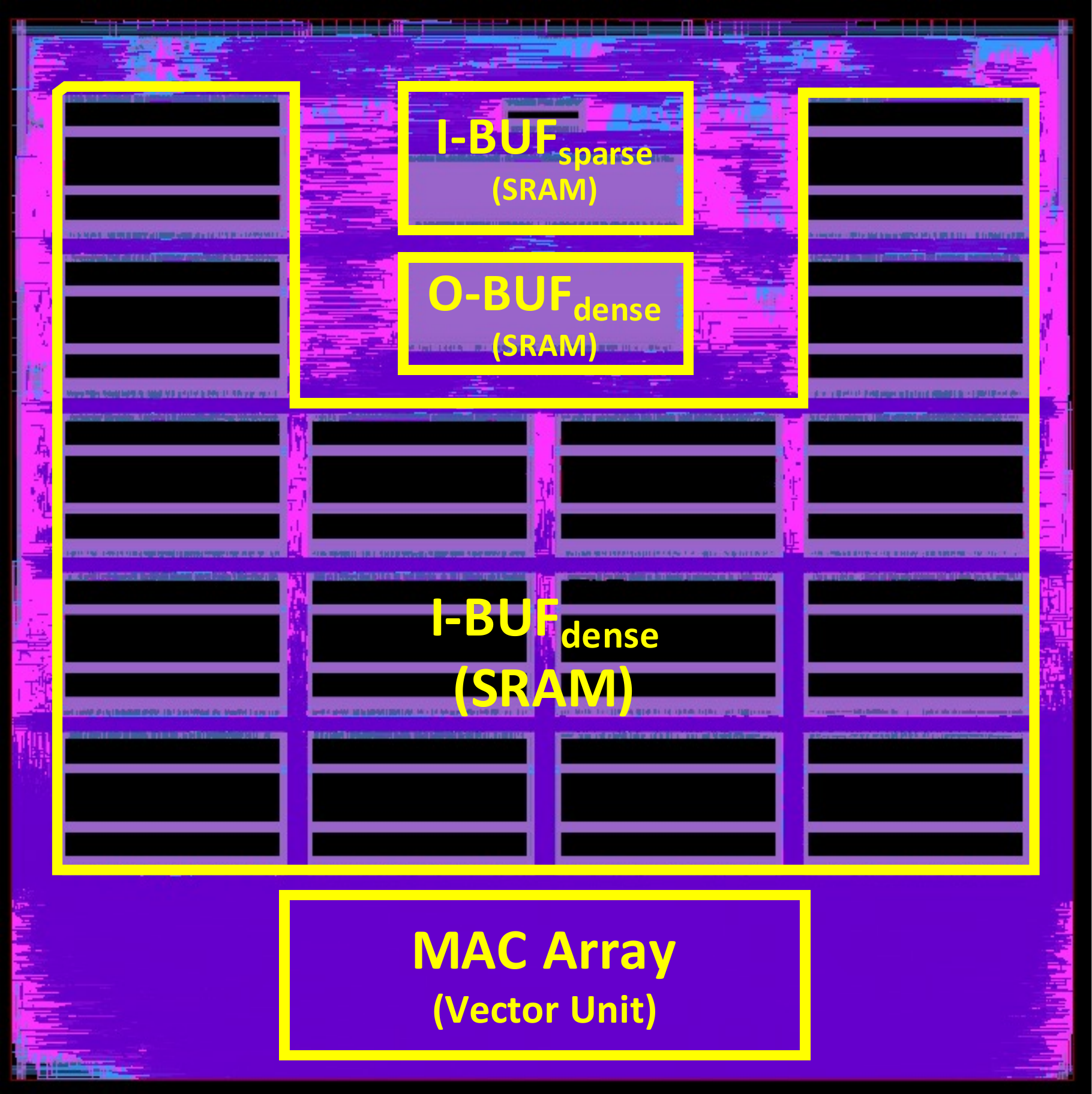}
\captionof{figure}{
	\proposed layout.
}
\label{fig:eval_die_photo}
\end{wrapfigure}
\par

The total area of \proposed is $5.8$\emph{mm$^{2}$} synthesized with a $65$ nm
standard-cell library (\tab{tab:grow_area}). The majority of area is used by
the on-chip SRAM buffers ($88\%$), especially the \ibufdense buffer (i.e., HDN
		cache and HDN ID list), which is a good design point to pursue as sparse
\spgemm algorithms are typically bottlenecked on off-chip memory bandwidth.

The area of GCNAX is reported as $6.5$\emph{mm$^{2}$} synthesized
under a $40$ nm technology. When \proposed's area is scaled down to $40$
nm, its area is estimated as $2.2$\emph{mm$^{2}$}. Given \proposed's superior
performance and energy-efficiency, \proposed provides an average $8.2\times$
better performance/mm$^{2}$ numbers. \fig{fig:eval_die_photo} shows the final
placed-and-routed design layout of GROW.

\subsection{Scalability}
\label{sect:eval_scalability}

\begin{figure}[t!] \centering
\includegraphics[width=0.485\textwidth]{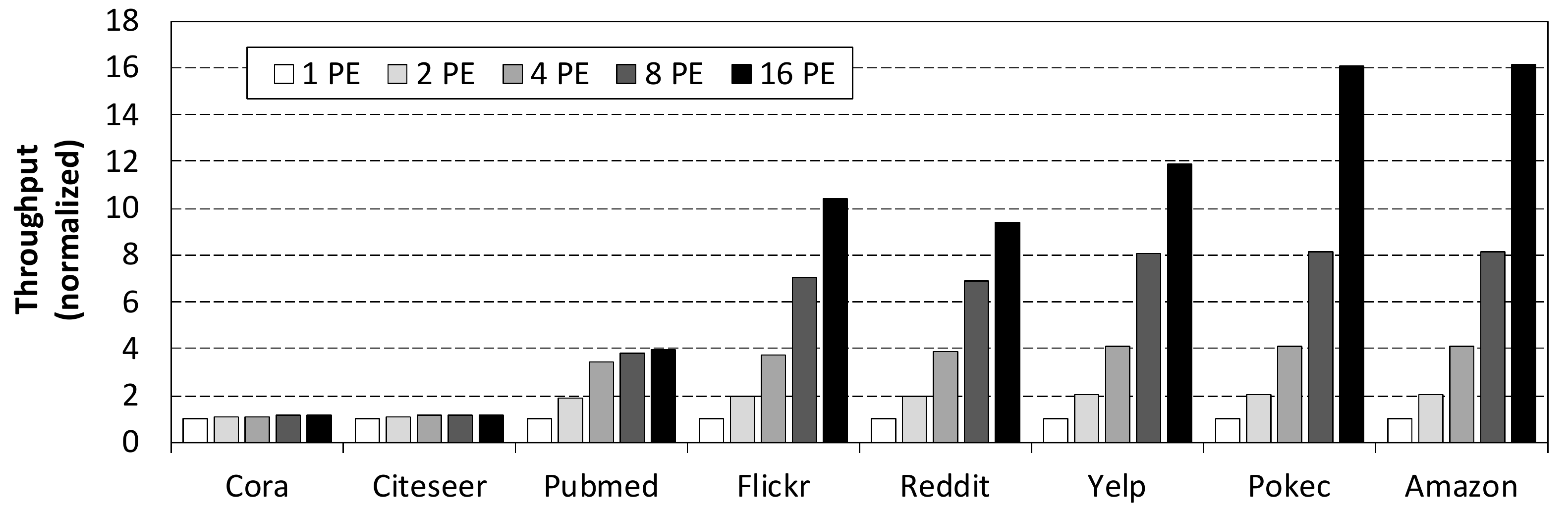}
\caption{
Performance scalability as a function of the number of PEs.
}
\vspace{-.2em}
\label{fig:eval_scalability}
\end{figure}

\fig{fig:eval_scalability} shows \proposed's performance when the number of
processing engines (PEs)  are swept from $1$ PE to
$16$ PEs with a proportional increase in memory bandwidth. For small-scale
graphs like Cora and Citeseer, having just a single \proposed PE is sufficient
to capture the entire working set, so the benefits of larger number of PEs are
small. As the input graph gets larger, however, the benefit of \proposed's
fine-grained row-stationary dataflow shines, achieving high scalability.  In
fact, for large-scale graphs like Yelp, Pokec and Amazon, \proposed achieves a
super-linear speedup.  
We observe that different PEs exhibit
	different memory intensive phases at different times as each PE handles a
		different graph cluster containing different loads. This leads to certain
		periods where a given PE is artificially given more than the average unit memory
		bandwidth (i.e., the memory bandwidth a single PE is provided with on average
				assuming perfect load balancing at steady state),
							which helps increase performance further.

\begin{figure}[t!] 
\centering
\subfloat[]{
\includegraphics[width=0.485\textwidth]{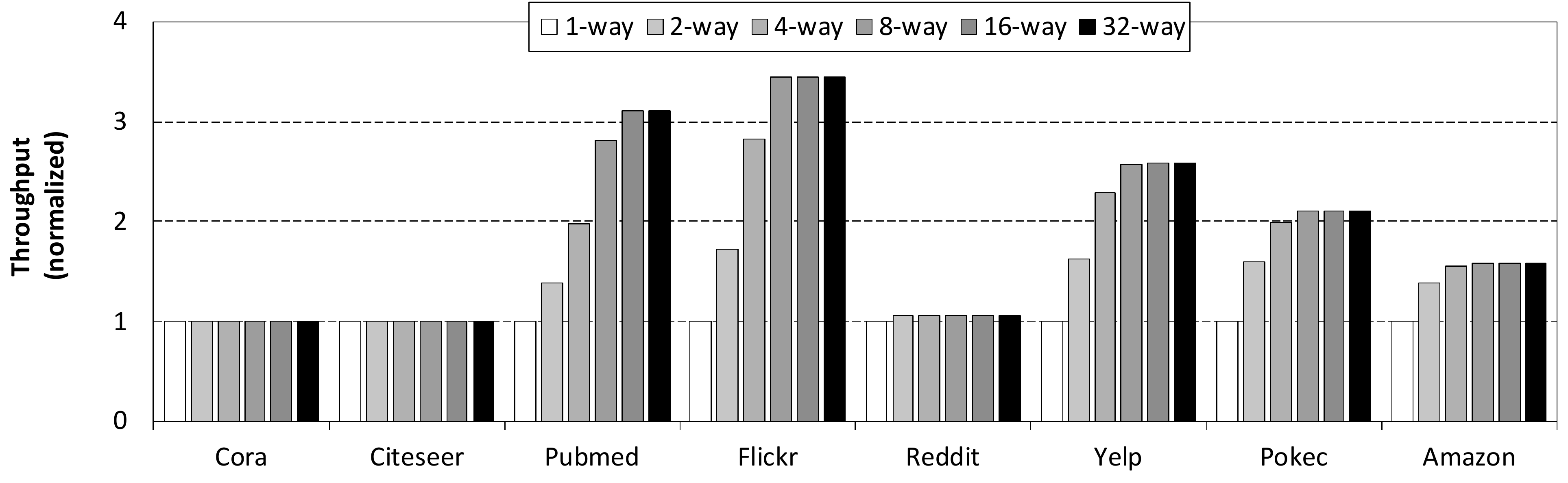}
\vspace{-0.2em}
	\label{fig:}
}
\vspace{0.3em}
\subfloat[]{
\includegraphics[width=0.485\textwidth]{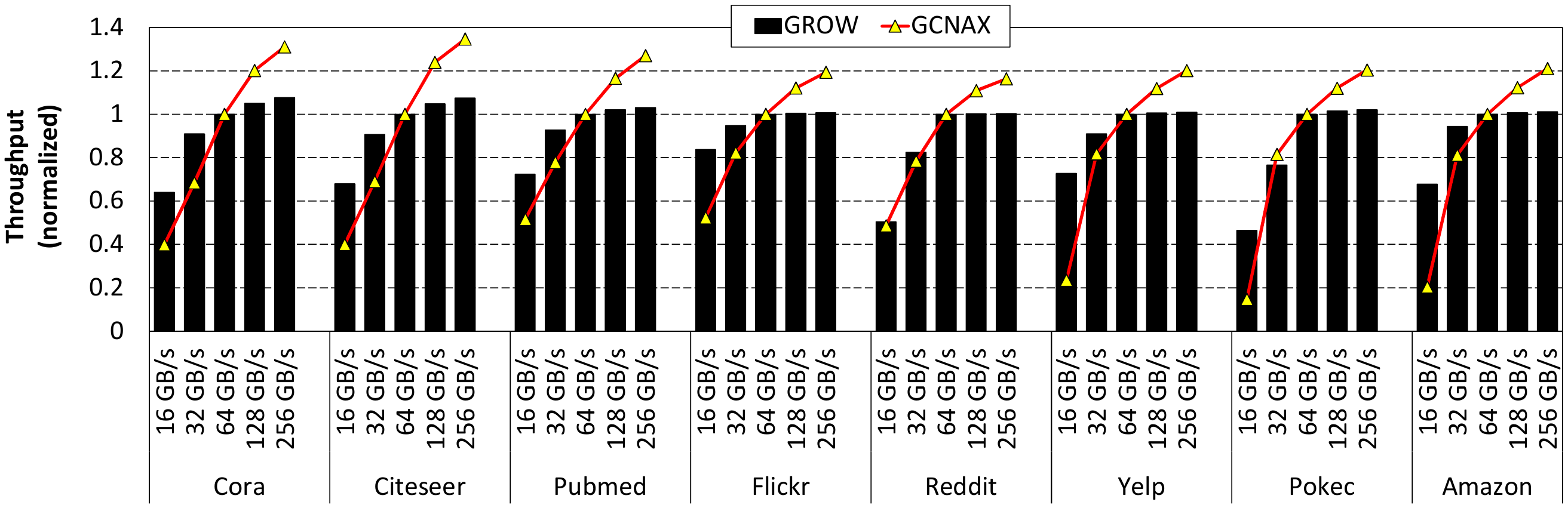}
	\vspace{-.2em}
	\label{fig:}
}
\caption{ 
\proposed sensitivity study. (a) Performance of \proposed when changing
	the degree of runahead execution. (b) Normalized performance of GCNAX and \proposed when changing the off-chip memory bandwidth, from $16$ GB/sec to $256$ GB/sec. In (b), note that both GCNAX and \proposed are each normalized to its own $64$ GB/sec design point in order to show the accelerator's sensitivity to memory bandwidth (i.e., the slope of GCNAX is much more steep than \proposed, meaning GCNAX suffers more from \spgemm's memory bandwidth limited characteristics).
}   
\vspace{-1.25em}
\label{fig:eval_sensitivity}
\end{figure}

\subsection{Sensitivity}
\label{sect:eval_sensitivity}

{\bf Runahead execution degree.} \fig{fig:eval_sensitivity}(a) shows
the performance of \proposed as we sweep the runahead execution degree
from $1$$-$$32$-ways. In five out of the eight workloads we study,
there is a sizable performance gap between $1$-way and $8$/$16$-way
runahead execution mode, demonstrating the importance of optimizing
\proposed memory-level parallelism using our multi-row stationary dataflow.

{\bf Off-chip memory bandwidth.} \fig{fig:eval_sensitivity}(b) shows the
changes in GCNAX and \proposed's throughput when the memory bandwidth
is swept from $16$ to $256$ GB/sec. Both GCNAX and \proposed are each
normalized to its own performance with $64$ GB/sec  memory bandwidth as means
	to highlight its robustness under different levels of memory
	throughput provided.  As depicted, GCNAX is highly sensitive to memory
	bandwidth showing a sharp increase (decrease) in performance when the
	provided memory bandwidth is increased (decreased).  In contrast, \proposed
	shows high robustness in performance thanks to its better utilization of
	memory bandwidth.

\subsection{Comparison to MatRaptor and GAMMA}
\label{sect:eval_gamma}

While the row-wise product based	 MatRaptor~\cite{matraptor} and
GAMMA~\cite{gamma}  
do not explore GCNs nor target
sparse-dense GEMM as in \proposed, given their adoption of Gustavson's
algorithm, we implement a cycle-level simulator of both MatRaptor/GAMMA to
demonstrate \proposed's merits. In \fig{fig:speedup_vs_gamma}, \proposed provides an
average $9.3\times$/$1.5\times$ speedup vs.
MatRaptor/GAMMA, respectively.  Reasons for \proposed's speedup are threefold.
First, MatRaptor/GAMMA are optimized for generic sparse-sparse GEMM so they 
fail to effectively exploit \emph{GCN-inherent} locality, i.e., MatRaptor does not employ
caches and GAMMA's fiber cache is not optimized for the power-law distribution
of graphs.  Second, sparse-sparse GEMM incurs a complicated and costly
partial-sum merging process, which is entirely redundant for \spgemm (i.e., key
		primitive of \proposed is a scalar $\times$ vector operation,
		\fig{fig:row_stationary}(b)), only to add performance and area overheads.
Third, the sparse-sparse GEMM assumes the RHS matrix is compressed in CSR which
adds additional indexing overheads as well as more memory traffic to fetch
metadata associated with CSR. On average, \proposed reduces memory traffic by
an average $18\times$/$4\times$ (maximum $46\times$/$7\times$) vs.
MatRaptor/GAMMA, respectively, leading to high speedup.

%% file: tex/discussion.tex
\section {Discussion} 
\label{sect:discussion}

\begin{figure}[t!] \centering
\includegraphics[width=0.495\textwidth]{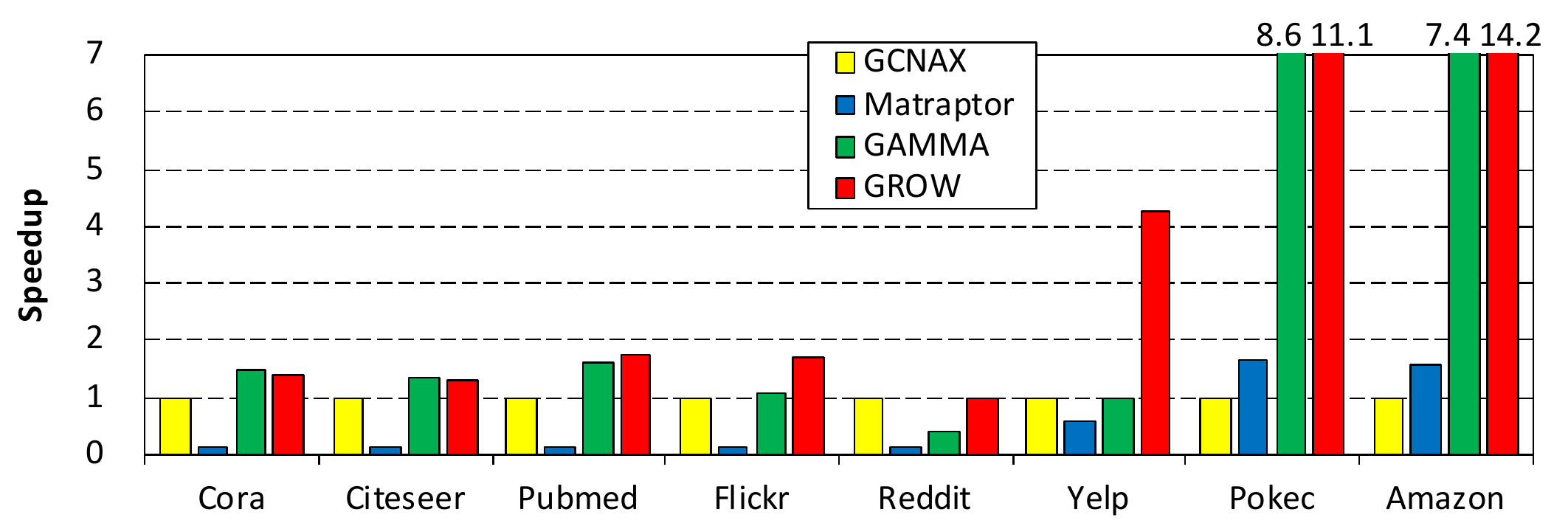}
\caption{
Performance comparison of \proposed vs. MatRaptor/GAMMA.
}
\vspace{-1em}
\label{fig:speedup_vs_gamma}
\end{figure}

{\bf GROW for non-power-law graphs.} 
	As the efficiency of graph partitioning is dependent upon the power-law
		distribution of input graphs, the effectiveness of GROW's HDN caching will 
		be reduced for non-power-law graphs.  Nonetheless, we expect the abundant
		parallelism reaped out by the combination of GROW's row-stationary dataflow
		and runahead execution to better hide latency than GCNAX, maintaining its
		superiority for even these challenging input datasets. Evaluation of GROW for such graphs however is beyond the scope of this paper and we leave it as future work.

{\bf Pinned vs. demand-based cache replacement policy.}
While GROW employs a policy where high-degree nodes are ``pinned''
inside the HDN cache, we also experimented with 
alternative cache eviction/replacement policies. For instance,  we also considered 
cache designs that seek to better
capture locality for low-degree nodes (e.g., 
the higher priority	high-degree node cache entries can get 
		evicted to store low-degree nodes per LRU policy) but
				it turns out that statically pinning
the high-degree nodes within the cache (as our current proposition) yielded the most
robust speedups. This is because the opportunity loss of evicting a high-degree node
greatly outweighed the benefits of capturing low-degree node's locality.

{\bf GROW applicability for advanced aggregation functions.}
Here we discuss GROW's applicability for more advanced 
aggregation functions discussed in prior work.
	 SAGEConv~\cite{pyg}
employs an aggregation function which conducts either a mean, pool, or LSTM operation over the \emph{sampled} neighbor nodes 
~\cite{graphsage}.
	Using the sampled node ID list, GROW's row-stationary
	dataflow can naturally fetch the sampled nodes from the $X$ matrix,
					 which then goes through mean, pool, or LSTM. Unlike the mean/LSTM
						 operators which GROW's existing microarchitecture (e.g., MAC array) can readily
						 be employed for derivation, 
						 pooling requires a separate, vector comparator array for accelerated computation.
When implemented and synthesized with $65$ nm standard-cell library, such comparator array
incurs $1.4\%$ additional area overhead vs. current GROW design.
GIN~\cite{gin} employs an aggregation function which adjusts the weight of the central node using a
	learnable parameter. As discussed in GCNAX, such aggregation function is refactored into multiple consecutive
	$W$ matrices so GROW is fully capable of supporting GIN as-is.
GAT~\cite{gat}
utilizes attention layers as its aggregation function, which involves MLP and softmax operators.
While MLPs can be computed using GROW's MAC array, derivation of softmax requires additional
microarchitectural support. There are multiple prior literature discussing 
hardware-accelerated softmax design (e.g., approximated polynomial based~\cite{poly}
vs. hash-table based~\cite{a3}),
	the most optimal design decision governed by the distribution of the values subject for softmax operation. Prior work on $A^3$~\cite{a3} reports that a high-overhead table-based softmax accelerator incurs approximately
		$16\%$ additional area overhead vs. its MAC array. When conservatively 
		projecting similar overheads
		to GROW's MAC array design, supporting GAT is estimated to incur a chip-wide $1.7\%$ area overhead. 
We leave the evaluation of GROW for these aggregation functions as future work as it is
beyond the scope of this paper.

%% file: tex/conclusion.tex
\section {Conclusion}
\label{sect:conclusion}

This paper presents a GCN inference accelerator named \proposed
based on the Gustavson's algorithm. 
	 \proposed is based on our software and hardware 
 co-design which effectively
	 balances locality and parallelism for high performance.
Unlike prior \spgemm based GCN accelerators, \proposed is
capable of intelligently exploiting the heterogeneous sparsity levels
manifested during the aggregation and combination phases of GCNs, drastically
reducing memory traffic for the memory-bound \spgemm algorithm.
Compared to state-of-the-art GCNAX, \proposed
reduces memory traffic by $2\times$, and achieves an average
$2.8\times$, $2.3\times$, and $8.2\times$ improvement in performance,
energy-efficiency, and performance/area, respectively.